\documentstyle[aps,epsf,rotate]{revtex}

\begin{document}

\title{Holes in the $t$-$J_z$ model: a thorough study}
\author{A. L. Chernyshev\cite{perm}}
\address{Physics Department, University of
California, Riverside, CA 92521}
\author{P. W. Leung}
\address{Physics Dept., Hong Kong University of Science and Technology,
Clear Water Bay, Hong Kong}
\date{\today}
\maketitle
\begin{abstract}
The $t$-$J_z$ model is the strongly anisotropic limit of the
$t$-$J$ model which captures some general properties
of the doped antiferromagnets (AF). The absence of spin
fluctuations simplifies the analytical  treatment of
hole motion in an AF background and  allows us to calculate the single- and
two-hole spectra with  high accuracy using regular diagram
technique combined with real-space approach.  
At the same time, numerical studies of this model via exact
diagonalization (ED) on small clusters show negligible finite
size effects for a number of quantities, 
thus allowing a direct comparison between  analytical
and numerical results. Both approaches demonstrate that the holes have
tendency to pair in the $p$-
and $d$-wave channels at realistic values of $t/J$.
The interactions leading to pairing and effects selecting $p$
and $d$ waves are thoroughly investigated.
The role of transverse spin fluctuations is considered using
perturbation theory. Based on the results of the present study,
we discuss the pairing problem in the
realistic $t$-$J$-like model.
Possible implications for preformed pairs formation and
phase separation are drawn.  
\end{abstract}
\section{Introduction}

The physics of holes moving in an AF background has received much
attention because of its possible connection to high-T$_c$
superconductivity. A microscopic realization of this physics is 
given by the $t$-$J$ model which
was first introduced as
a conjecture \cite{And}, and then derived as a low-energy limit of a
realistic model representing the electronic structure of cuprates
\cite{ZR,Bel}. 
Recent comparison of the $t$-$J$ model results to the
angle-resolved photoemission data \cite{ARPES} showed that the overall shape of
the experimental single-hole band can be fitted satisfactorily using the
pure $t$-$J$ model. To account for the detail
line shapes, one needs to include more distant hopping terms
($t^{\prime}$, $t^{\prime\prime}$, etc.)
\cite{Naz,Bel1}.  The
presence of these terms also follows from a
careful mapping study \cite{Bel1}. 
Numerical studies of the $t$-$J$ model by means of ED and other methods
suggest the presence of hole binding in the physical parameter range
$t/J\sim 3$ \cite{Dag,Dag1,Ed}, and that the dominant symmetry of the
pairing correlation is $d_{x^2-y^2}$. This finding also supports the
common belief that some variant of the $t$-$J$ model is able to describe the
physics of the real compounds.
However, to show that this $d$-wave binding in the $t$-$J$ model is relevant
to the physics of high-$T_c$ materials, one has to 
clarify how strong and generic the reasons behind this
pairing are.
Unfortunately, numerical studies alone are of limited use for this
purpose, so that an analytical study is necessary
to develop an insight into the problem.

In this paper we attempt such a study by investigating the Ising limit
of the $t$-$J$ model --
the $t$-$J_z$ model. It is well-known to be a simplified limiting case of the
$t$-$J$ model, but only recently  has its properties
been fairly well understood. In the pioneering work by Bulaevskii {\it
et al} \cite{BNK} the single-hole energy in the $t/J\ll 1$ limit has
been calculated. Subsequently, Brinkmann and Rice \cite{BR}
considered the $J=0$ limit and introduced the retraceable-path approximation.
More recent works \cite{KLR,SVR,HM,Zaan,Reiter} used
the self-consistent Born approximation (SCBA) to study the $t$-$J_z$,
$t$-$J$, and  $t$-$t^\prime$-$J$ models. Using this approximation,
an analytical expression for
the hole Green's function in the $t$-$J_z$ model for arbitrary
values of $t/J$ has been found by Starykh and Reiter \cite{Star}.
The pairing of holes in  the $t$-$J_z$ model has been considered in the works
by Trugman \cite{Trug1} and by Shraiman and Siggia \cite{SS}, where
important results were obtained. Some other studies used different
modifications of the variational approach \cite{Ed1,CDS} and obtained
similar results. However, as we shall show, previous analytical
treatments involved approximation which prevented one from
obtaining the correct answers. 
 
The $t$-$J_z$ model has also been studied using numerical methods \cite{Dag}.
Barnes {\it et al}
\cite{Barnes} compared their ED results on a 16-site square
lattice with some analytical results, 
and Riera and Dagotto \cite{RD} developed a
modified Lanczos technique which enable them to study
the one- and two-hole
ground states of the model on lattices as large as 50-sites. 
(For a review see Ref. \cite{Dag}). It was shown that
finite-size effects in the ED data of the $t$-$J_z$ model
are much smaller than those of the $t$-$J$ model \cite{RD}.

In this paper we develop an analytical treatment of the $t$-$J_z$ model
based on the results of Ref. \cite{Star}.
Comparison with our ED data allows us to justify the validity of this 
analytical approach. 
Furthermore, we use the results of this study to shed light
on the problem of whether the pairing in the $d$-channel is a generic
feature of $t$-$J$-like models, and what interaction defines the
symmetry of the ground state of the system.

Our numerical ED results for the $t$-$J_z$ model are obtained on a
32-site square lattice using 
the same method as in Refs.~\cite{lg95,clg98}.
This is up to now the only ED results on the largest
lattice.
This numerically exact method allows us to analyze
various properties of the single-hole ground state  as well as
the two-hole bound states with different symmetries.

Our paper is organized as follows. Sections II and III 
describe the representation and Feynman rules used in this
paper. In Sec. IV we solve the single-hole problem and compare the
results to ED data. In Sec. V the two-hole problem is considered in
detail. In Sec. VI we summarize our results and draw conclusions.

\section{Representations of the Hamiltonian and operators}

The Hamiltonian of the $t$-$J_z$ model is
\begin{eqnarray}
\label{H}
{\cal H} = -t\sum_{\langle ij\rangle\sigma}(\tilde{c}^\dagger_{i\sigma}
\tilde{c}_{j\sigma}+{\rm H.c.})+ 
J\sum_{\langle i,j \rangle} \bigl[S_i^z   S_j^z-
\frac{1}{4}N_iN_j \bigr] 
\end{eqnarray}
where the summation runs over nearest-neighbor bonds
$\langle ij\rangle$,
$\tilde{c}^\dagger_{i\sigma}=c^\dagger_{i\sigma}(1-n_{i\bar{\sigma}})$,
$\tilde{c}_{i\sigma}=(\tilde{c}^\dagger_{i\sigma})^\dagger$,
$N_i=\tilde{c}^\dagger_{i\uparrow}\tilde{c}_{i\uparrow}
+\tilde{c}^\dagger_{i\downarrow}\tilde{c}_{i\downarrow}$.
Since we are modeling
electrons on a 2D square lattice, the coordination number $z=4$ and the
spin $S=\frac{1}{2}$.

Assuming the presence of long-range antiferromagnetic order one can introduce
a representation for the Hubbard operators (operators of spin
and constrained fermion) through the spinless fermion and
boson operators. The choice of the representation is, in general,
arbitrary and is motivated by the problem to be solved. 
In our work we use a generalization of the Dyson-Maleev (DM)
representation for the Hubbard operators.
As usual, such a representation conserves the algebra of the
operators but extends the Hilbert space of the problem. 

In sublattice $A=\{\uparrow\}$ the DM representation is given by
\begin{eqnarray}
\label{rep1}
&&S_i^z=\frac{1}{2}\bigl(X_i^{\uparrow\uparrow}-
X_i^{\downarrow\downarrow}\bigr)=\frac{1}{2}-a_i^{\dag}a_i-
\frac{1}{2}f_i^{\dag}f_i\ , \ \ \ \ 
S_i^-=X_i^{\downarrow\uparrow}=a_i^{\dag}\bigl(1-a_i^{\dag}a_i-
f_i^{\dag}f_i\bigr)\ , \ \ \ \ S_i^{\dag}=X_i^{\uparrow\downarrow}=a_i
\nonumber\\
&&\tilde{c}_{i\uparrow}=X_i^{0\uparrow}=
f_i^{\dag}\bigl(1-a_i^{\dag}a_i\bigr)\ , \ \ \ \
\tilde{c}^\dagger_{i\uparrow}=X_i^{\uparrow 0}=f_i\ , \ \ \ \
\tilde{c}_{i\downarrow}=X_i^{0\downarrow}=f_i^{\dag}a_i\ , \ \ \ \
\tilde{c}^\dagger_{i\downarrow}=X_i^{\downarrow 0}=f_i a_i^{\dag}\ , \ \ \ \
\\
&&N_i=\tilde{c}^\dagger_{i\uparrow}\tilde{c}_{i\uparrow}+
\tilde{c}^\dagger_{i\downarrow}\tilde{c}_{i\downarrow}=
X_i^{\uparrow\uparrow}+X_i^{\downarrow\downarrow}=1-X_i^{00}=
1-f_i^{\dag}f_i\ .
\nonumber
\end{eqnarray}
In sublattice $B=\{\downarrow\}$, the corresponding representation is given by
\begin{eqnarray}
\label{rep2}
&&S_j^z=\frac{1}{2}\bigl(X_j^{\uparrow\uparrow}-
X_j^{\downarrow\downarrow}\bigr)=-\frac{1}{2}+b_j^{\dag}b_j+
\frac{1}{2}g_j^{\dag}g_j\ , \ \ \ \ 
S_j^-=X_j^{\downarrow\uparrow}=b_j\ , \ \ \ \ 
S_j^{\dag}=X_j^{\uparrow\downarrow}=b_j^{\dag}\bigl(1-b_j^{\dag}b_j-
g_j^{\dag}g_j\bigr)\ ,
\nonumber\\
&&\tilde{c}_{j\uparrow}=X_j^{0\uparrow}=g_j^{\dag}b_j\ , \ \ \ \
\tilde{c}^\dagger_{j\uparrow}=X_j^{\uparrow 0}=g_j b_j^{\dag}\ , \ \ \ \
\tilde{c}_{j\downarrow}=X_j^{0\downarrow}=g_j^{\dag}\bigl(1-
b_j^{\dag}b_j\bigr)\ , \ \ \ \
\tilde{c}^\dagger_{j\downarrow}=X_j^{\downarrow 0}=g_j\ , \ \ \ \
\\
&&\tilde{c}^\dagger_{j\uparrow}\tilde{c}_{j\uparrow}+
\tilde{c}^\dagger_{j\downarrow}\tilde{c}_{j\downarrow}=
X_j^{\uparrow\uparrow}+X_j^{\downarrow\downarrow}=1-X_j^{00}=
1-g_j^{\dag}g_j\ .
\nonumber
\end{eqnarray}
Thus, we have two types of fermions and bosons associated with the $A$ and
$B$ sublattices.
The Hamiltonian in Eq.(\ref{H}) expressed in terms of the new variables
Eqs.(\ref{rep1},\ref{rep2}) is
\begin{eqnarray}
\label{H1}
&&{\cal H}\Rightarrow{\cal H}_0+{\cal H}_1+{\cal H}_2 \ , 
\nonumber\\
&&{\cal H}_0 = \frac{1}{2}
J\sum_{\langle i,j \rangle} \bigl[a_i^{\dag}a_i+b_j^{\dag}b_j
+f_i^{\dag}f_i+g_j^{\dag}g_j\bigr]\ ,\nonumber\\
&&{\cal H}_1 = t\sum_{\langle ij\rangle}\bigl[f_i^{\dag}g_j
(b_j^{\dag}+a_i)+{\rm H.c.}\bigr]\ ,\\
&&{\cal H}_2=-t\sum_{\langle ij\rangle}\bigl[f_i^{\dag}g_j
b_j^{\dag}a_i^{\dag}a_i+g_j^{\dag}f_ia_i^{\dag}b_j^{\dag}b_j
\bigr]-
\frac{1}{2}J\sum_{\langle i,j \rangle}\bigl[
n_i^f b_j^{\dag}b_j +n_j^g a_i^{\dag}a_i
+n_i^f n_j^g +2a_i^{\dag}a_ib_j^{\dag}b_j\bigr]
\ .
\nonumber
\end{eqnarray}
Here we have divided 
the Hamiltonian
Eq.(\ref{H1})
into three parts: ${\cal H}_0$, consisting of the linear spin-excitation
and hole terms, ${\cal H}_1$ consisting of the bare hole-magnon interaction
term, and ${\cal H}_2$, consisting of the 
nonlinear hole-magnon, static hole-magnon, direct hole-hole, 
and magnon-magnon interaction terms.
The third and fourth terms in ${\cal H}_0$ take
into account the energy of the four AF bonds
broken by a hole introduced into the system. 
In the following this energy is included in the ground state energy
and the corresponding terms are omitted.
Note that we have omitted a constant term. 
The same Hamiltonian Eq.(\ref{H1}) in ${\bf k}$-space is given by
\begin{eqnarray}
\label{H2}
&&{\cal H}_0 = 2J\sum_{\bf q}\bigl[
a_{\bf q}^{\dag}a_{\bf q}+b_{\bf q}^{\dag}b_{\bf q}\bigr]+
2J\sum_{\bf k}\bigl[
f_{\bf k}^{\dag}f_{\bf k}+g_{\bf k}^{\dag}g_{\bf k}\bigr]\ , 
\nonumber\\
&&{\cal H}_1 = 4t\sum_{\bf k,q}  
\gamma_{\bf k-q}\bigl[f_{\bf k-q}^{\dag}g_{\bf k}b_{\bf q}^{\dag}
+g_{\bf k-q}^{\dag}f_{\bf k}a^{\dag}_{\bf q}
+{\rm H.c.}\bigr]\ ,
\\
&&{\cal H}_2=-4t\sum_{\bf k,q,q_1,q_2}\gamma_{\bf k-q}
\bigl[f_{\bf k-q-q_1+q_2}^{\dag}g_{\bf k}b_{\bf q}^{\dag}
a_{\bf q_1}^{\dag}a_{\bf q_2}+g_{\bf k-q-q_1+q_2}^{\dag}f_{\bf k}
a_{\bf q}^{\dag} b_{\bf q_1}^{\dag}b_{\bf q_2}\bigr]
\\
&&\phantom{{\cal H}_1=}-
2J\sum_{\bf k,q,q_1}\gamma_{\bf q-q_1}\bigl[
f_{\bf k-q+q_1}^{\dag}f_{\bf k} b_{\bf q}^{\dag}b_{\bf q_1} 
+g_{\bf k-q+q_1}^{\dag}g_{\bf k} a_{\bf q}^{\dag}a_{\bf q_1} \bigr]
\nonumber\\
&&\phantom{{\cal H}_1=}
-2J\sum_{\bf k,k^{\prime},q}\gamma_{\bf q}
f_{\bf k-q}^{\dag}g_{\bf k^{\prime}+q}^{\dag}g_{\bf k^{\prime}}f_{\bf k}
-4J\sum_{\bf q_1,q_2,q}\gamma_{\bf q}
a_{\bf q_1-q}^{\dag}b_{\bf q_2+q}^{\dag}b_{\bf q_2}a_{\bf q_1}\ ,
\nonumber
\end{eqnarray}
where $\gamma_{\bf k}=(\cos(k_x)+\cos(k_y))/2$, and all summations are
restricted to the first magnetic Brillouin zone (BZ).

\section{propagators and vertex functions}

The Feynman rules for the model in Eq. (\ref{H2}) are:\\
(i) Hole propagator: \ \ 
\begin{minipage}[t]{7.cm}
$G^0(\epsilon)=
G_{f(g)}^0(\epsilon)=[\epsilon+i0]^{-1}$
\end{minipage}\hskip 1.2cm
\begin{minipage}[t]{6.cm}
\begin{figure}
\unitlength 1cm
\epsfxsize=0.9cm
\begin{picture}(0,0.5)
\put(0,0.5){\rotate[r]{\epsffile{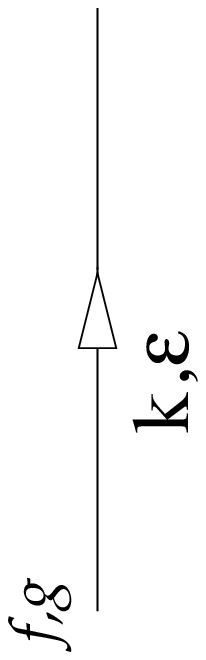}}}
\end{picture}
\end{figure}
\end{minipage}\\
(ii) Magnon propagator: \ \ 
\begin{minipage}[t]{7.cm}
$D^0(\omega)=D_{a(b)}^0(\omega)
=[\omega-2J+i0]^{-1}$
\end{minipage}\hskip 0.5cm
\begin{minipage}[t]{6.cm}
\begin{figure}
\unitlength 1cm
\epsfxsize=1.3cm
\begin{picture}(1,0.5)
\put(0,0.2){\rotate[r]{\epsffile{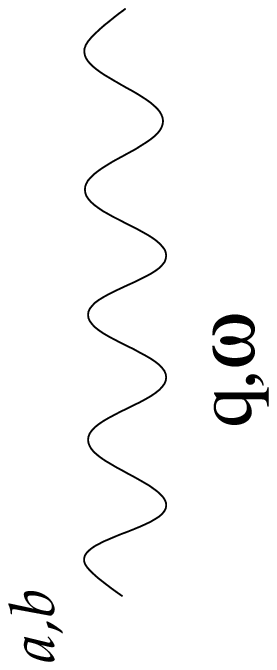}}}
\end{picture}
\end{figure}
\end{minipage}\\
(iii) Hole-magnon vertex:\\
\begin{minipage}[t]{7.5cm}
\ \\

 \hskip 0.7 cm $\Gamma^0(k,q,k-q)=4t\gamma_{\bf k-q}\cdot
\delta_{\epsilon_1,\epsilon-\omega}$
\end{minipage}\hskip 1cm
\begin{minipage}[t]{6.cm}
\begin{figure}
\unitlength 1cm
\epsfxsize=2cm
\begin{picture}(2,2)
\put(0,0.7){\rotate[r]{\epsffile{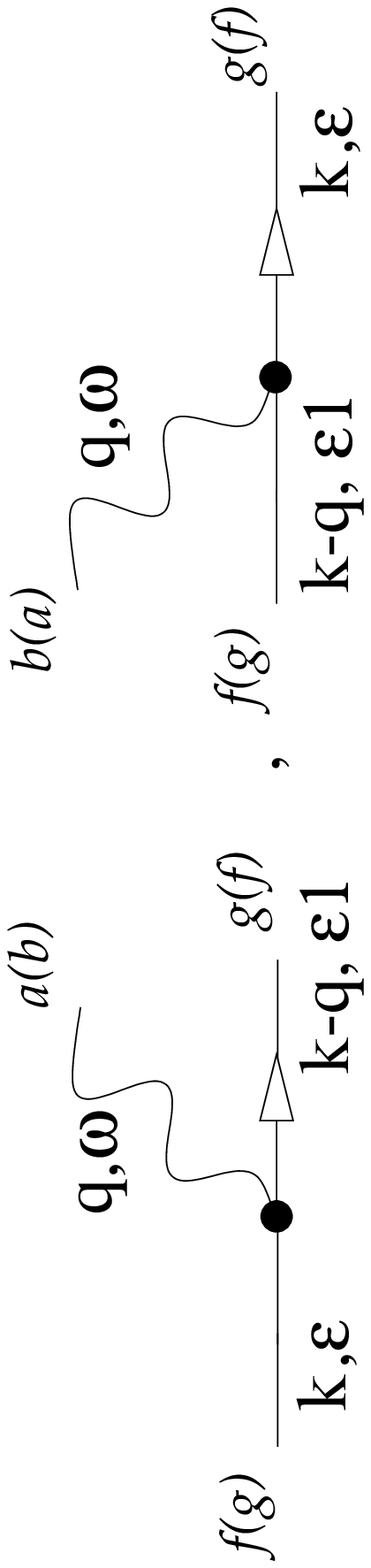}}}
\end{picture}
\end{figure}
\end{minipage}\\
The indices indicate that $f$ (or $g$)-holes can emit only 
$a$ (or $b$)-magnons and absorb 
only $b$ (or $a$)-magnons. This is an important feature of the hole-magnon
interaction, and is a consequence of spin conservation.
In the following we omit the indices.
We use the following shorthand notations ---
$\delta_{\epsilon_1,\epsilon-\omega}$ denotes
$\delta(\epsilon_1,\epsilon-\omega)$ and
$\sum_\omega$ means $\int_{-\infty}^{\infty}\frac{d\omega}{2\pi i}$. \\
(iv) Hole-two-magnon vertex:\\
\begin{minipage}[t]{9.cm}
\ \\

 \hskip 0.7 cm $\Gamma_2^0(k,q,q_1,k-q+q_1)=
-2J\gamma_{\bf q-q_1}\cdot\delta_{\epsilon_1,\epsilon-\omega+\omega_1}$
\end{minipage}\hskip 2cm
\begin{minipage}[t]{6.cm}
\begin{figure}
\unitlength 1cm
\epsfxsize=2cm
\begin{picture}(2,2)
\put(0,0.7){\rotate[r]{\epsffile{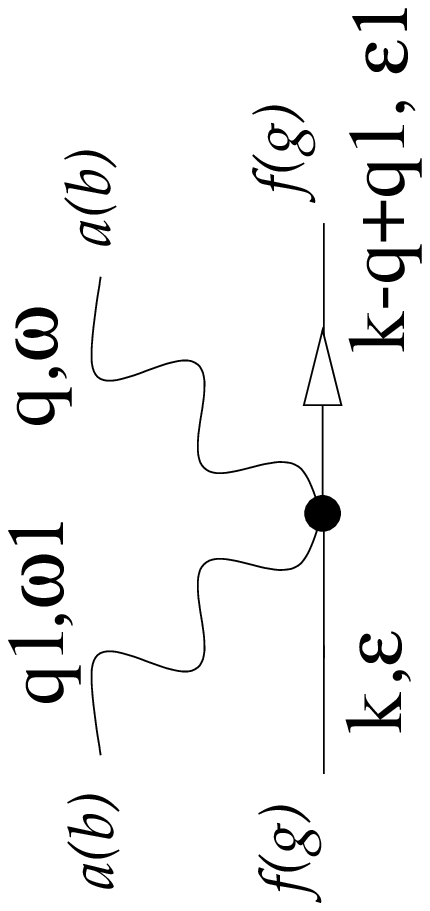}}}
\end{picture}
\end{figure}
\end{minipage}\\
Note that $f$ (or $g$)-holes interact with $a$ (or $b$)-magnons only.\\
(v) Hole-three-magnon vertex:\\
\begin{minipage}[t]{9.cm}
\ \\

 \hskip 0.7 cm $\Gamma_3^0(k,q,q_1,q_2,k-q-q_1+q_2)=
-4t\gamma_{\bf k-q}\cdot\delta_{\epsilon_1,\epsilon-\omega-
\omega_1+\omega_2}$
\end{minipage}\hskip 2cm
\begin{minipage}[t]{6.cm}
\begin{figure}
\unitlength 1cm
\epsfxsize=2.5cm
\begin{picture}(2,2)
\put(0,0.5){\rotate[r]{\epsffile{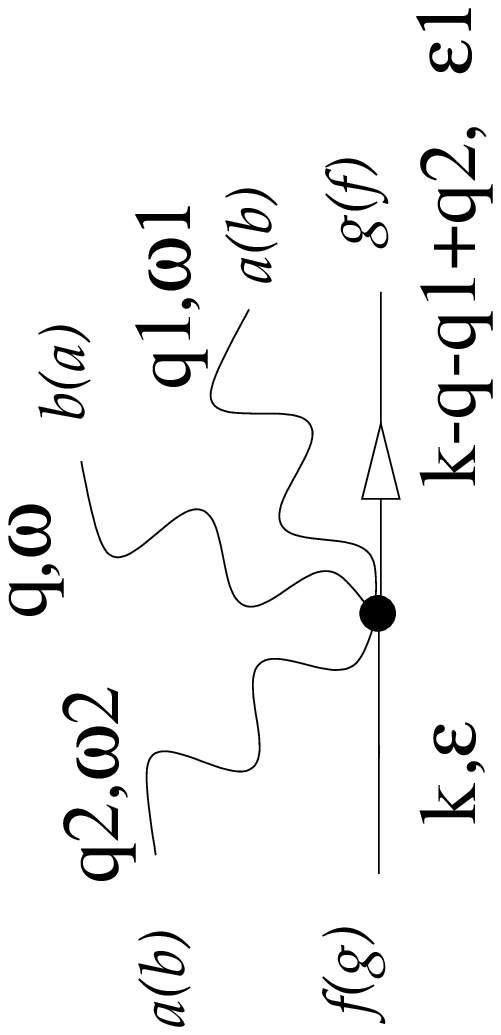}}}
\end{picture}
\end{figure}
\end{minipage}\\
Note that there is no vertex diagram for the reversed order of
emission-absorption because of the non-hermitian nature of the
corresponding terms in the Hamiltonian Eq.(\ref{H2}).\\
(vi) Hole-hole vertex:\\
\begin{minipage}[t]{9.cm}
\ \\

 \hskip 0.7 cm $\Gamma_{fg}^0(k,k^{\prime},q)=-2J\gamma_{\bf q}\cdot
\delta_{\epsilon_1+\epsilon_2,\epsilon_3+\epsilon_4}$
\end{minipage}\hskip 2cm
\begin{minipage}[t]{6.cm}
\begin{figure}
\unitlength 1cm
\epsfxsize=2cm
\begin{picture}(2,2)
\put(0,0.7){\rotate[r]{\epsffile{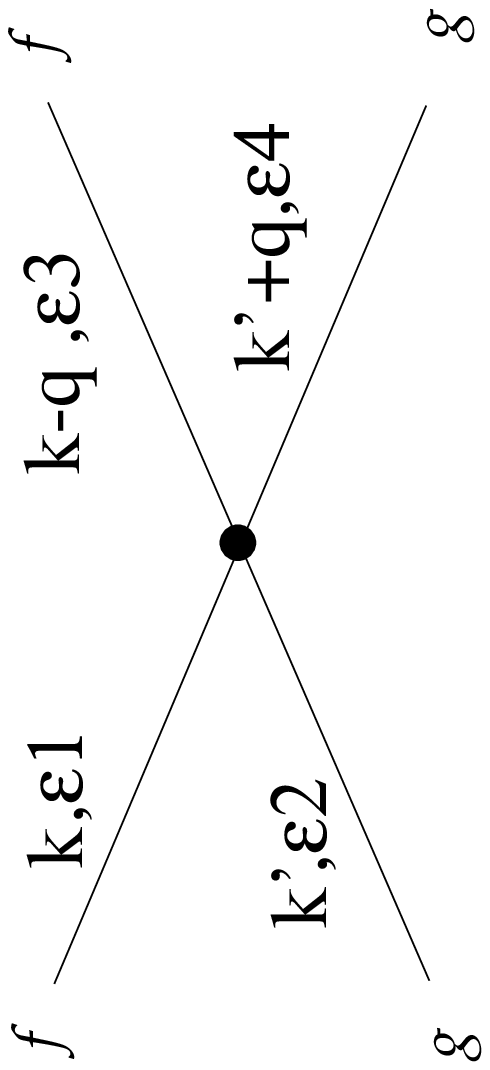}}}
\end{picture}
\end{figure}
\end{minipage}\\
(vii) Magnon-magnon vertex:\\
\begin{minipage}[t]{9.cm}
\ \\

 \hskip 0.7 cm $\Gamma_{ab}^0(q_1,q_2,q)=-4J\gamma_{\bf q}\cdot
\delta_{\omega_1+\omega_2,\omega_3+\omega_4}$
\end{minipage}\hskip 2cm
\begin{minipage}[t]{6.cm}
\begin{figure}
\unitlength 1cm
\epsfxsize=2cm
\begin{picture}(2,2)
\put(0,0.7){\rotate[r]{\epsffile{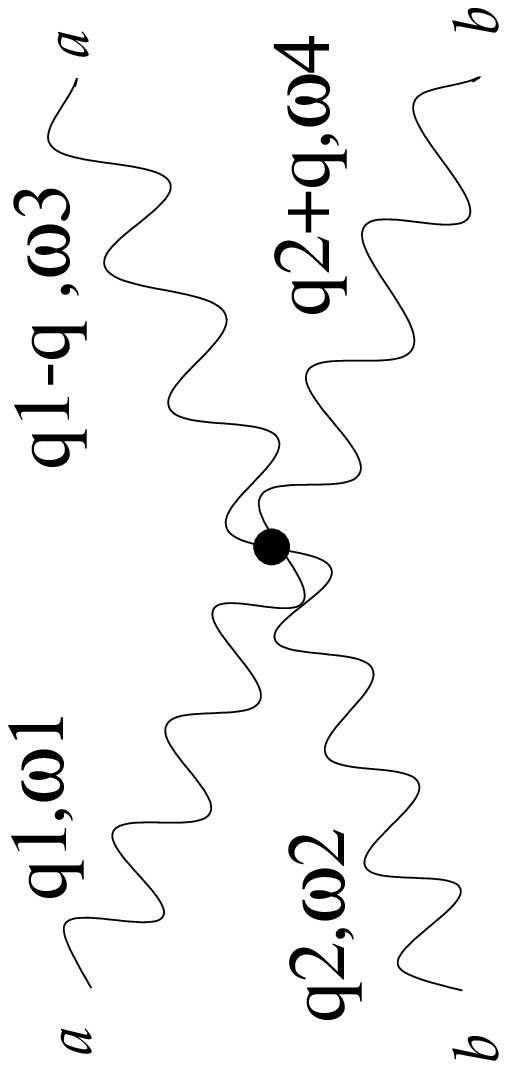}}}
\end{picture}
\end{figure}
\end{minipage}\\
where we use the notations $k=({\bf k},\epsilon)$ and 
$q=({\bf q},\omega)$ for the momentum and frequency of the hole and 
magnon respectively.

Thus the model Eq.(\ref{H2}) involves fermion-fermion (hole-hole), 
boson-boson (magnon-magnon), and three types of fermion-boson
(hole-magnon) interactions. In general the problem of finding the low-energy
excitations in such a model is very complicated. However, as we will
show below, in 
the case of the $t$-$J_z$ model it is possible to take into
account the result of renormalization almost exactly using a
regular diagrammatic treatment.

\section{Single hole}

\begin{figure}
\unitlength 1cm
\epsfxsize=1.5cm
\begin{picture}(7,3)
\put(1,1){\rotate[r]{\epsffile{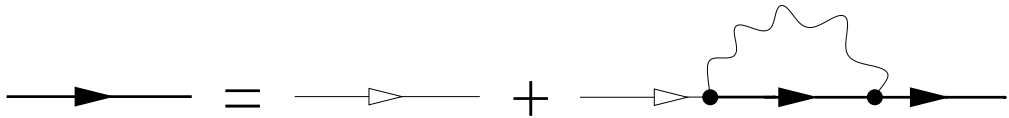}}}
\end{picture}
\caption{SCBA for the hole self-energy. Bold lines represent the dressed
  Green's functions.}
\label{diag1}
\end{figure}

If we ignore ${\cal H}_2$, the
Hamiltonian in Eq.~(\ref{H2}) for a system with a single-hole
is
similar to the problem of a single electron interacting with the
local phonon mode. When the interaction ($t$ in our case) is small
compared to the energy of the boson ($2J$), we
can use perturbation theory and consider
only the lowest contribution to the hole energy, which
corresponds to the virtual emission-absorption of a single magnon.
At larger $t$ many-magnon intermediate states must be considered.
Such a problem, if solved self-consistently, must take
into account 
the renormalization of the fermion-boson interaction, or crossing
diagrams.
However, in the $t$-$J$ model it is well known that the
lowest correction to the bare hole-magnon vertex $\Gamma^0$ is
exactly zero since the hole-magnon interaction conserves the spin,
or more exactly, the pseudospin associated with the sublattice type.
Therefore, the crossing diagrams can be omitted and the
renormalization of the hole energy is given only 
by the series of diagrams
where the magnons are absorbed in exactly the reversed order
in which they are emitted.
This is the so-called self-consistent
Born approximation (SCBA), which in the case of the $t$-$J_z$ model
is identical to the ``retraceable path approximation''. This
coincidence is due to the local nature of the spin-excitations in the
model. The corresponding diagrammatic equation for the hole Green's
function is shown in
Fig. \ref{diag1}. Note that the vertices are unrenormalized.
\noindent
The retraceable path approximation implies that the hole is confined by
a ``string'' of spin excitations and the hole motion is 
completely incoherent, i.e. the hole form a localized state.

\begin{figure}
\unitlength 1cm
\epsfxsize=2cm
\begin{picture}(2,3)
\put(7,0.5){\rotate[r]{\epsffile{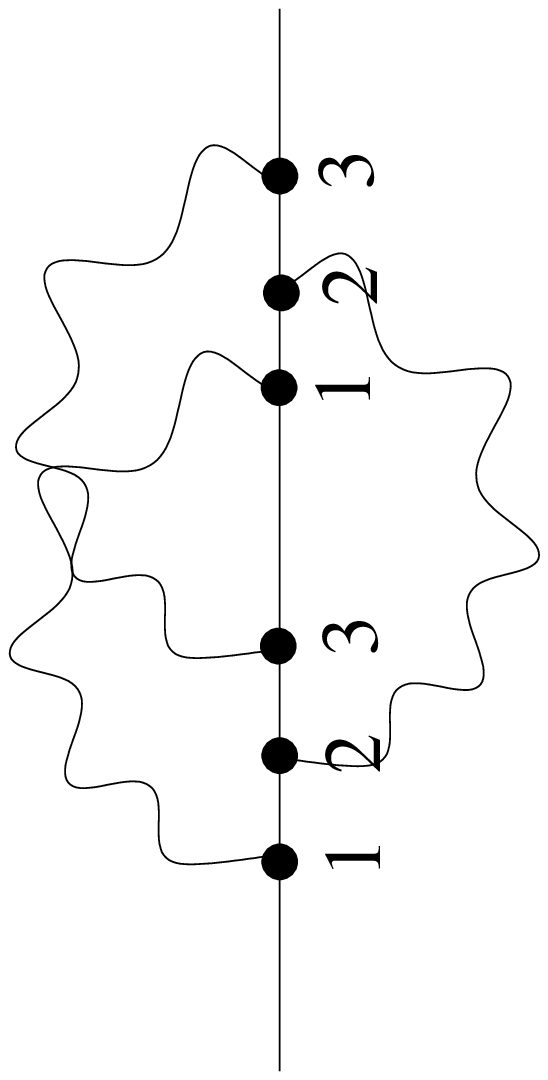}}}
\end{picture}
\caption{Trugman's diagram. The numbers represent the sites where a magnon is
emitted (absorbed).}
\label{trug}
\end{figure}

The first non-zero contribution to the hole self-energy  
from the crossing diagrams is shown in Fig. \ref{trug}. It is
of sixth order in $t$ and 
is small compared to the contribution of the ``retraceable'' 
diagrams of the same (sixth) order due to the geometrical factor. 
\noindent
This diagram is the first member of the family of the so-called ``Trugman
paths''. Their contribution to the hole energy is small 
in a wide range of $t/J$, but they are responsible for the 
finite coherent band of the hole. Real-space
consideration shows that such a diagram corresponds to the hole's
motion around an elementary square
loop in which the hole makes one and a half turn. 
As a result all spin excitations are cured and the hole is
translated along the diagonal of the square plaquette, i.e. to a next-nearest
neighbor. 
As we have noted, the corresponding correction to the energy 
is small and we will neglect 
these diagrams in the rest of this Section.

Since $G^0(\epsilon)$ and $D^0(\omega)$ are
momentum-independent and the hole-magnon interaction depends only on
${\bf k-q}$, the full Green's function $G$ is also ${\bf k}$-independent. 
Thus the equation shown in Fig. \ref{diag1} can be simplified as
\begin{eqnarray}
\label{SCBA}
&&\Sigma(\epsilon)=\sum_{\omega}G(\epsilon-\omega)D^0(\omega)
\sum_{\bf q} (\Gamma^0_{\bf k-q})^2=4t^2G(\epsilon-2J) \ ,
\nonumber\\
&&G(\epsilon)=[\epsilon-4t^2G(\epsilon-2J)]^{-1}\ .
\end{eqnarray}
The solution of Eq.~(\ref{SCBA}) at all $t/J$ has been found in
Ref. \cite{Star}. One can try an ansatz for $G$ of the form:
\begin{eqnarray}
\label{anz}
G(\epsilon)=-\frac{1}{2t}\frac{Y(\epsilon)} 
{Y(\epsilon+2J)} \ .
\end{eqnarray}
This transforms Eq.~(\ref{SCBA}) into a difference equation:
\begin{eqnarray}
\label{anz1}
Y(\epsilon+2J)+Y(\epsilon-2J)=-\frac{\epsilon}{2t}Y(\epsilon) \ ,
\end{eqnarray}
which is the recursion relation for the Bessel functions. Thus
$G(\epsilon)$ is given by
\begin{eqnarray}
\label{G1}
G(\epsilon)=-\frac{1}{2t}\frac{J_{-\epsilon/2J}(2t/J)} 
{J_{-\epsilon/2J-1}(2t/J)}\ ,
\end{eqnarray}
where $J_\nu(x)$ are the Bessel functions.

Such a form is the consequence of the continued fraction form of the
Green's function, which in turn is the result of 
the retraceable path
approximation. Note that in Ref. \cite{Star} $J_z=2J$.
Thus the poles of the Green's function in the SCBA approximation 
are defined by the zeros of the Bessel function. Since all these zeros are
real, the hole spectral function consists of the set of delta-function
peaks corresponding to the energy levels of the 
quasiparticle states in the ``string'' potential well. 
Inclusion of the Trugman processes does not
change this result significantly because the resulting 
bandwidth is much smaller than the separation of the levels.

Although the solution (\ref{G1}) of Eq.~(\ref{SCBA}) is an almost exact 
solution of  ${\cal H}_0+{\cal H}_1$, it is not a solution
of the original $t$-$J_z$ Hamiltonian. Besides the crossing diagrams,
there are other contributions to the self-energy originating from
${\cal H}_2$. In order to compare
with other analytical approaches and ED, it is necessary to account for
these contributions. 

To make this statement evident let us now consider the 
two limiting cases $t/J\ll 1$ and
$t/J\gg 1$, and  
compare the results of the SCBA with the known facts from 
the real-space approach and ED numerical data.

For $t/J\ll 1$ the ground-state energy for 
the hole is given by Eqs. (\ref{SCBA}) as
\begin{eqnarray}
\label{w_0}
\epsilon_0^{SCBA}=-\frac{2t^2}{J}\ .
\end{eqnarray}
This is inconsistent with the result of the real-space approach
\begin{eqnarray}
\label{w_0_1}
\epsilon_0=-\frac{8t^2}{3J}\ .
\end{eqnarray}
The reason for this discrepancy is that the spin-flip (magnon) has
lower energy ($3J/2$ compared to $2J$) 
if it is located in the neighborhood of the hole, and the
Hamiltonian ${\cal H}_0+{\cal H}_1$ does not take this difference into account.
Note that throughout this paper we assign zero energy level to the state
with one {\it static} hole, i.e. $E_0=E_0^{Ising}+2J=0$.

According to Ref. \cite{Star}, in the large $t$ limit 
the ground-state energy given by Eq. (\ref{G1}) is
\begin{eqnarray}
\label{w_0_2}
\epsilon_0^{SCBA}=-4t+2\beta_0 t(J/t)^{2/3}-2J\ ,
\end{eqnarray}
where $\beta_0=2.34$ is the first zero of the Airy function. 
There is no alternative asymptotic expression in this limit which
is accurate up to order $J$, but the first
two terms which are of order
$t$ and $t^{1/3}J^{2/3}$ can be compared to the
results of the ``string'' approach by Shraiman and Siggia (SS) \cite{SS} and 
the fitting of 16-site ED data by Barnes {\it et al} \cite{Barnes}
\begin{eqnarray}
\label{w_0_3}
&&\epsilon_0^{SS}=-2\sqrt{3} t+2.74 t(J/t)^{2/3}\ ,\\
&&\epsilon_0^{ED}=-3.63 t+2.93 t(J/t)^{2/3}\ .
\nonumber
\end{eqnarray}
This comparison makes one to suspect that  SCBA (Eq.(\ref{w_0_2})) 
overestimates both the absolute value
of the depth of the ``band'' and the slope of the walls of the linear
confining potential. 

\begin{figure}
\unitlength 1cm
\epsfxsize=2cm
\begin{picture}(2,3)
\put(3,0.5){\rotate[r]{\epsffile{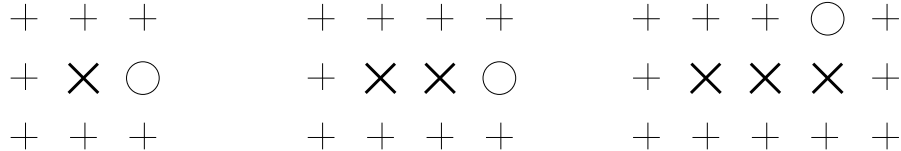}}}
\end{picture}
\caption{One-, two-, and three-magnon strings}
\label{mov}
\end{figure}

The origin of these discrepancies becomes evident when we consider the
real-space picture of the retraceable hole movement. Fig. \ref{mov}
shows the real-space images of three components of the spin-polaron
wave function: the one-, two-, and three-magnon strings. The first
important observation is that the coordination number in the creation
of the first magnon is indeed $z=4$, but is $z-1=3$ in the
creation of each next magnon.
This is why the factor in the
first term of the SS expression Eq.~(\ref{w_0_3}) for the hole
energy is $2\sqrt{3}$ (instead of $4$). If $t$ is much larger than $J$, the 
average length of the strings contributing to the spin-polaron wave
function is large. Then the weight of the components with 
coordination number different from others
 becomes insignificant. Since this component is
the ``bare'' hole, it also means that the quasiparticle residue of the
state decreases as $t$ grows and in the $t\rightarrow\infty$ limit the
spectrum will be incoherent. This result was first obtained
by Brinkmann and Rice \cite{BR} in the $J=0$ limit.
Having in mind
that the energy of spin excitation is determined by the number of
broken AF bonds associated with its creation, one can see that the
energy of the first magnon in the string is $3J/2$ (3 bonds, $J/2$ each), 
but the energy of each subsequent magnon is just $J$ (2 bonds). 
This
is true for all strings of length $l=2$ and for most other
longer strings except for those which have self-tangencies. 
Since the latter cases are relatively rare due to the
geometrical factor, one can assign the same energy $l\cdot J+J/2$ to every
string of  length $l$.
A more accurate account for the energy of the self-tangient strings 
makes only tiny changes to the
energy of the whole excitation. Thus, making the changes $2J\rightarrow J$ 
and $z\rightarrow z-1$ in
the expression for the SCBA hole energy will change 
the factor $2\beta_0=4.68$ in
the second term of Eq.~(\ref{w_0_2}) to
$3^{1/6}\beta_0=2.81$, leaving it
in reasonable agreement with Eq. (\ref{w_0_3}). 

Summarizing the results of the comparison one can say that the reason for
the failure of the SCBA is that ${\cal H}_0+{\cal H}_1$ 
is not the same as 
${\cal H}_{t-J}$ and that an improved
formalism must take into account the following facts: (i) the energy
of the first magnon in a string is $3J/2$, (ii) the energy of the $n$th
magnon ($n>1$) is $J$, (iii) the coordination number is $z-1=3$ for 
$l\geq 1$ and $z=4$ for $l=0$, where $l$ is the length of the string.
After some thinking using the above real-space arguments one can
suggest the following modification to
the equation for the Green's function (\ref{SCBA}) to meet 
all these requirements
\begin{eqnarray}
\label{G2}
&&G(\epsilon)=[\epsilon-4t^2G_a(\epsilon-3J/2)]^{-1}\ ,\nonumber
\\
&&G_a(\epsilon)=[\epsilon-3t^2G_a(\epsilon-J)]^{-1}\ ,\\
&&G_a(\epsilon)=-\frac{1}{\sqrt{3}t}\frac{J_{-\epsilon/J}(2\sqrt{3}t/J)} 
{J_{-\epsilon/J-1}(2\sqrt{3}t/J)}\ .
\nonumber
\end{eqnarray}
Such a modification
gives the correct hole energy in the small $t$
limit.  In the $t\rightarrow\infty$ limit
the resulting hole energy is $-2\sqrt{3}t$. 

Let us now come back to the original problem with the full Hamiltonian 
${\cal H}_0+{\cal H}_1+{\cal H}_2$, Eq. (\ref{H2}) and discuss how such an
``improvement'' of the Green's function proposed in Eq.(\ref{G2}) can be done
formally in the diagrammatic language. 

\begin{figure}
\unitlength 1cm
\epsfxsize=2cm
\begin{picture}(3,3)
\put(2,0.5){\rotate[r]{\epsffile{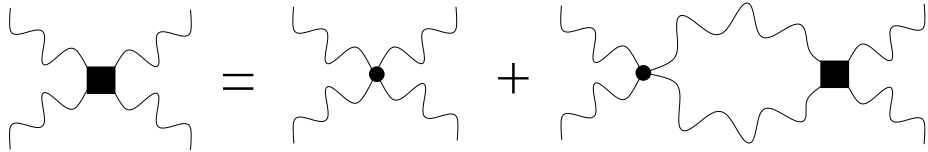}}}
\end{picture}
\caption{The Bethe-Salpeter equation for the two-magnon problem. 
Dots represent
bare vertices $\Gamma_{ab}^0$, and squares represent full vertices
$\Gamma_{ab}$.}
\label{diag2}
\end{figure}

The fact that magnons in the string have lower energy
than free excitations is the result of
magnon-magnon binding. The magnon-magnon interaction (last term of
${\cal H}_2$, Eq. (\ref{H2})) is attractive and leads to a bound
state. A regular approach to the bound state problem is based on the
Bethe-Salpeter equation shown in Fig. \ref{diag2}.
\noindent
The equivalent integral equation is
\begin{eqnarray}
\label{BSm}
&&\Gamma_{ab}(\Omega,{\bf q_1-q_3})=\Gamma^0_{ab}({\bf q_1-q_3})+
\sum_{\bf p_1}\Gamma^0_{ab}({\bf q_1-p_1})\Gamma_{ab}(\Omega,{\bf p_1-q_3})
\sum_\omega D^0(\omega)D^0(\Omega-\omega)
\\
&&\phantom{\Gamma_{ab}(\Omega,{\bf q_1-q_3})}
=-4J\gamma_{\bf q_1-q_3}+\sum_{\bf p_1}
(-4J\gamma_{\bf q_1-p_1})\frac{\Gamma_{ab}(\Omega,{\bf p_1-q_3})}
{\Omega-4J+i0}
\ .
\nonumber
\end{eqnarray}
It is easy to see that the momentum and frequency dependence of the
vertex can be separated. The suggestion $\Gamma_{ab}(\Omega,{\bf q_1-q_3})
=-4J\gamma_{\bf q_1-q_3}\tilde{\Gamma}_{ab}(\Omega)$ transforms
Eq. (\ref{BSm}) into
\begin{eqnarray}
\label{BSm1}
\tilde{\Gamma}_{ab}(\Omega)=1-\frac{J}{\Omega-4J+i0}
\tilde{\Gamma}_{ab}(\Omega)\ .
\end{eqnarray}
Here we have used the relation $\sum_{\bf p_1}\gamma_{\bf q_1-p_1}
\gamma_{\bf p_1-q_3}=\gamma_{\bf q_1-q_3}/4$. Thus the renormalized
magnon-magnon vertex is given by 
\begin{eqnarray}
\label{BSm2}
\Gamma_{ab}(\Omega,{\bf k-p})=-4J\gamma_{\bf k-p}
\frac{\Omega-4J+i0}{\Omega-3J+i0}\ .
\end{eqnarray}
The pole of this function, $\Omega=3J$, corresponds to the bound state
whose energy is lower than that of two free magnons by $-J$,
in agreement with our expectations.

\begin{figure}
\unitlength 1cm
\epsfxsize=4.5cm
\begin{picture}(6,5)
\put(0,0.5){\rotate[r]{\epsffile{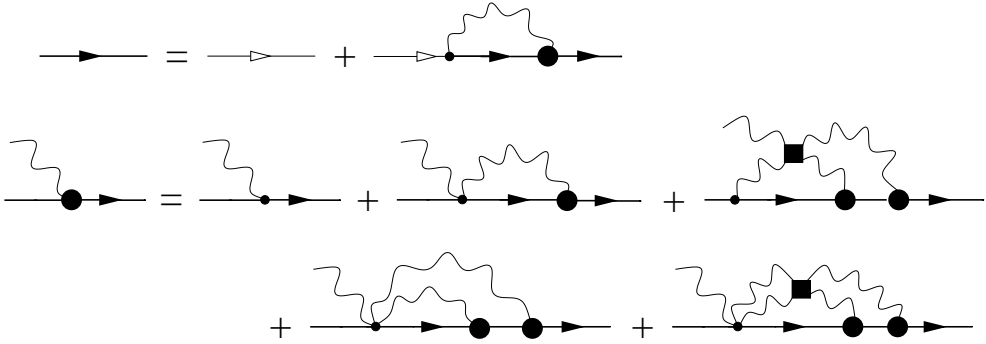}}}
\end{picture}
\caption{The Dyson equation for the hole Green's function. 
Bold lines are the dressed Green's
  functions, circles are the renormalized hole-magnon vertices, squares
  are the renormalized magnon-magnon vertices (Fig. \ref{diag2} and
  Eq. (\ref{BSm2})).}
\label{diag3}
\end{figure}

Similarly the hole-two-magnon interaction (second term 
in ${\cal H}_2$, Eq. (\ref{H2}))
lowers the energy of the magnons in the neighborhood of a
hole. The hole-three-magnon interaction (first term in ${\cal H}_2$) 
originates from the projection operator and 
projects out the unphysical states with the hole and magnon at
the same site created by the hopping term in ${\cal H}_0$.
A consistent consideration of these interactions yields the following
Dyson equation for the single-hole propagator:
\begin{eqnarray}
\label{Dyson}
&&G(\epsilon)=[\epsilon-\Sigma(\epsilon)]^{-1} \ ,
\nonumber\\
&&\Sigma(\epsilon)=\sum_{\omega}G(\epsilon-\omega)D^0(\omega)
\sum_{\bf q} \Gamma^0_{\bf k-q} \Gamma_{\bf k,q}(\epsilon,\omega)\ ,
\end{eqnarray}
where the renormalized hole-magnon vertex is 
\begin{eqnarray}
\label{Dyson1}
&&\Gamma_{\bf k,q}(\epsilon,\omega)=\Gamma^0_{\bf k-q}+
\Gamma_{\bf k,q}^{(1)}(\epsilon)+
\Gamma_{\bf k,q}^{(2)}(\epsilon,\omega)
+\Gamma_{\bf k,q}^{(3)}(\epsilon)
+\Gamma_{\bf k,q}^{(4)}(\epsilon)\ ,
\\
&&\ \ \ \ \Gamma_{\bf k,q}^{(1)}(\epsilon)=
\sum_{{\bf q_1},\omega_1}\Gamma_2^0({\bf q-q_1})G(\epsilon-\omega_1)
D^0(\omega_1)\Gamma_{\bf k,q_1}(\epsilon,\omega_1)\ ,
\nonumber\\
&&\ \ \ \ \Gamma_{\bf k,q}^{(2)}(\epsilon,\omega)=
\sum_{\bf q_1,q_2}\sum_{\omega_2,\Omega}
\Gamma_{\bf k-q-q_1}^0 D^0(\Omega-\omega)G(\epsilon-\Omega)
\Gamma_{ab}(\Omega,{\bf k-q_2})D^0(\omega_2)D^0(\Omega-\omega_2)
G(\epsilon-\omega_2)
\nonumber\\
&&\ \ \ \ \phantom{\Gamma_{\bf k,q}^{(2)}(\epsilon,\omega)=
\sum_{\bf q_1,q_2}\sum_{\omega_2,\Omega}}
\times\Gamma_{\bf k,q+q_1}(\epsilon-\omega_2,\Omega-\omega_2)
\Gamma_{\bf k,q_2}(\epsilon,\omega_2)\ ,
\nonumber\\
&&\ \ \ \ \Gamma_{\bf k,q}^{(3)}(\epsilon)=
\sum_{\bf q_1,q_2}\sum_{\omega_1,\omega_2}
\Gamma_3^0({\bf k-q-q_2})D^0(\omega_1)D^0(\omega_2)
G(\epsilon-\omega_1-\omega_2)
\Gamma_{\bf k-q-q_2}(\epsilon-\omega_1,\omega_2)
G(\epsilon-\omega_1)
\Gamma_{\bf k,q_1}(\epsilon,\omega_1)\ ,
\nonumber\\
&&\ \ \ \ \Gamma_{\bf k,q}^{(4)}(\epsilon)=
\sum_{\bf q_1,q_2,Q}\sum_{\omega_1,\omega_2,\Omega}
\Gamma_3^0({\bf k-q-Q})D^0(\omega_1)D^0(\Omega-\omega_1)
G(\epsilon-\Omega)\Gamma_{ab}(\Omega,{\bf q_1-q_2})
D^0(\omega_2)D^0(\Omega-\omega_2)
\nonumber\\
&&\ \ \ \ \phantom{\Gamma_{\bf k,q}^{(4)}(\epsilon)=
\sum_{\bf q_1,q_2,Q}\sum_{\omega_1,\omega_2,\Omega}}
\times\Gamma_{\bf k,q_1+Q}(\epsilon-\omega_2,\Omega-\omega_2)
G(\epsilon-\omega_2)
\Gamma_{\bf k,q_2}(\epsilon,\omega_2)\ .
\nonumber
\end{eqnarray}
A diagrammatic representation of Eqs. (\ref{Dyson},\ref{Dyson1}) 
is shown in Fig. \ref{diag3}.

Note that the lowest order correction to {\it all} vertices are
zero, and higher order corrections are neglected. 
We have already discussed the absence of
one-loop correction to the hole-magnon vertex.
It is less evident in cases of the hole-two- and hole-three-magnon
vertices. However, if one recalls the $f,g$ and $a,b$ indices for the
hole and magnon lines in Fig. \ref{diag3}, it becomes evident that the
emission of an extra magnon prior to the hole-two-magnon interaction
alters the type of the hole and therefore prevents it from absorbing and
reemitting the original magnon. The same is true for the hole-three-magnon
vertex.
It is interesting to note that the set of diagrams in Fig. \ref{diag3}
is still in the SCBA, i.e., the crossing diagrams
are absent. 

Again the momentum and frequency dependence of $\Gamma$ can be separated.
The almost evident substitution 
$\Gamma_{\bf k,q}(\epsilon,\omega)=4t\gamma_{\bf k-q}
\tilde{\Gamma}(\epsilon,\omega)$ simplifies
Eqs. (\ref{Dyson},\ref{Dyson1}) which, 
after integrating over the internal frequencies and momenta, 
becomes
\begin{eqnarray}
\label{Dyson2}
&&\Sigma(\epsilon)=4t^2
G(\epsilon-2J)\tilde{\Gamma}(\epsilon,2J)\ ,
\nonumber\\
&&\tilde{\Gamma}(\epsilon,\omega)=1-
G(\epsilon-2J)\tilde{\Gamma}(\epsilon,2J)
\biggl[\frac{J}{2}-4t^2\frac{J}{\omega-J}
\biggl(G(\epsilon-3J)\tilde{\Gamma}(\epsilon-2J,J)-
G(\epsilon-2J-\omega)\tilde{\Gamma}(\epsilon-2J,\omega)\biggr)
\\
&&\phantom{\tilde{\Gamma}(\epsilon,\omega)=1}
+t^2 G(\epsilon-4J)\tilde{\Gamma}(\epsilon-2J,2J)
+t^2 \biggl(G(\epsilon-3J)\tilde{\Gamma}(\epsilon-2J,J)-
G(\epsilon-4J)\tilde{\Gamma}(\epsilon-2J,2J)\biggr)\biggl]\ .
\nonumber
\end{eqnarray}
Each term in the square bracket comes from the corresponding
irreducible diagram in Fig. \ref{diag3}. After combining similar terms
$\tilde{\Gamma}(\epsilon,2J)$ becomes
\begin{eqnarray}
\label{Dyson3}
&&\tilde{\Gamma}(\epsilon,2J)=G(\epsilon-2J)^{-1}
\biggl[G(\epsilon-2J)^{-1}+J/2
-3t^2G(\epsilon-3J)\tilde{\Gamma}(\epsilon-2J,J)+
4t^2 G(\epsilon-4J)\tilde{\Gamma}(\epsilon-2J,2J)\biggr]^{-1}
\\
&&\phantom{\tilde{\Gamma}(\epsilon,2J)}
=G(\epsilon-2J)^{-1}
\biggl[\epsilon-3J/2-3t^2G(\epsilon-3J)\tilde{\Gamma}(\epsilon-2J,J)
\biggr]^{-1} =\frac{G_a(\epsilon-3J/2)}{G(\epsilon-2J)}\ ,
\nonumber
\end{eqnarray}
where $G_a(\epsilon)$ has a continued fraction form. Summarizing
Eqs. (\ref{Dyson}),(\ref{Dyson2}),(\ref{Dyson3}) one finally obtains
\begin{eqnarray}
\label{Dyson4}
&&G(\epsilon)=[\epsilon-4t^2G_a(\epsilon-3J/2)]^{-1}\ ,
\\
&&G_a(\epsilon)=[\epsilon-3t^2G_a(\epsilon-J)]^{-1}
=-\frac{1}{\sqrt{3}t}\frac{J_{-\epsilon/J}(2\sqrt{3}t/J)} 
{J_{-\epsilon/J-1}(2\sqrt{3}t/J)}\ ,
\nonumber
\end{eqnarray}
which is exactly the same as the Green's function 
suggested in Eq. (\ref{G2}).

\begin{figure}
\unitlength 1cm
\epsfxsize=9cm
\begin{picture}(9,7.5)
\put(5,-4){\epsffile{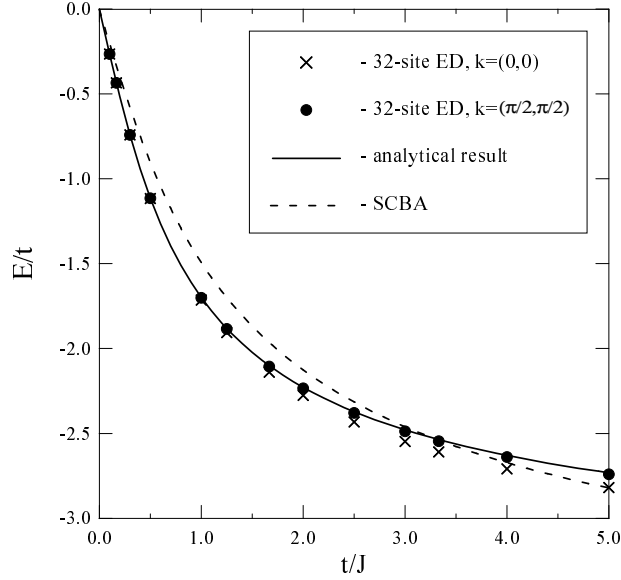}}
\end{picture}
\caption{The single-hole ground state energy v.s. $t/J$.
The zero level of energy is taken to be the energy of 
the hole at $t=0$. 
}
\label{comp1}
\end{figure}

The rest of this Section is devoted to the comparison 
of the results from Eq. (\ref{Dyson4}) with our numerical ED results
on a 32-site lattice.
Fig. \ref{comp1} shows the ED results together
with analytical results from the SCBA Eq. (\ref{SCBA})
and from Eq. (\ref{Dyson4}).
 
\noindent
We recall that contributions from the Trugman's paths
are left out from Eqs.~(\ref{Dyson4}) and (\ref{SCBA}).
These paths give rise to a small dispersion of the hole,
$\delta E_{\bf k}=t_1^{eff}\cos(k_x)\cos(k_y)+
t_2^{eff}(\cos(2k_x)+\cos(2k_x))+\dots$, with $t_2^{eff}\ll t_1^{eff}$.
This form of dispersion
implies that the correction to the energy at the point $(\pi/2,\pi/2)$ 
due to the Trugman paths is almost zero.
Therefore, it is natural
to compare the ED data at this ${\bf k}$-point to the analytical results.
From Fig.~\ref{comp1} one can see a beautiful agreement 
of the numerical data at
${\bf k}=(\pi/2,\pi/2)$ with the results of Eq. (\ref{Dyson4})
in the whole range of $t/J$.
For comparison purpose, Fig.~\ref{comp1} also shows the numerical
data at ${\bf k}=(0,0)$. 
We also remark that our 32-sites ED-data at ${\bf k}=(0,0)$
agree with the 50-sites results of the modified Lanczos study 
by Riera and Dagotto \cite{RD} (not shown in Fig. \ref{comp1})
up to
the fourth digit at $t/J=5$ and up to the seventh digit at $t/J=1.25$. 

\begin{figure}
\unitlength 1cm
\epsfxsize=9cm
\begin{picture}(9,8.5)
\put(5,-4){\epsffile{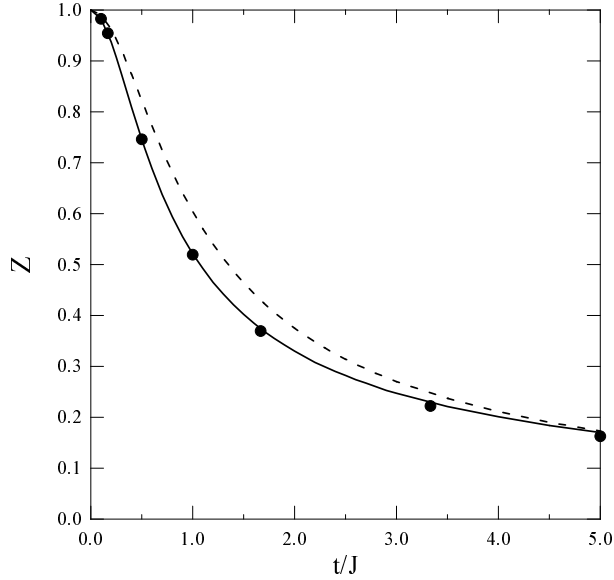}}
\end{picture}
\caption{Quasiparticle residue for the lowest pole v.s. $t/J$. 
Dashed curve is the result of the SCBA. Solid curve is the result of
the present approach. Dots are the numerical data for the
$(\pi/2,\pi/2)$ point.}
\label{comp2}
\end{figure}

It is already clear from Fig.~\ref{comp1} that our present approach
works better than the SCBA. To further illustrate this point,
we plot the quasiparticle residue in
Figure \ref{comp2}. Three sets of data are shown. They are the results
from our present analytical approach, the SCBA, and our numerical ED study.
The ED results are calculated using the same
method as in Ref.~\cite{lg95}.
One can see that such a comparison unambiguously favor the present
approach. 

\section{Two holes}

In this section we consider
the problem of hole pairing in the $t$-$J_z$ model
using the diagrammatic formalism introduced in previous
sections. 
Our goal is to study in detail the bound state of two holes. 
The role of different interactions and higher order corrections
will be analyzed. Bound states with $s$-, $p$-, and $d$-symmetry
and zero total momentum will be investigated. 
Throughout this paper we consider the $S^z=0$ bound states only.
Consequently we are interested in the 
interactions of the particles originating from different
sublattices ($f$ and 
$g$ fermions). 

\begin{figure}
\unitlength 1cm
\epsfxsize=2.5cm
\begin{picture}(3,3)
\put(1,0.5){\rotate[r]{\epsffile{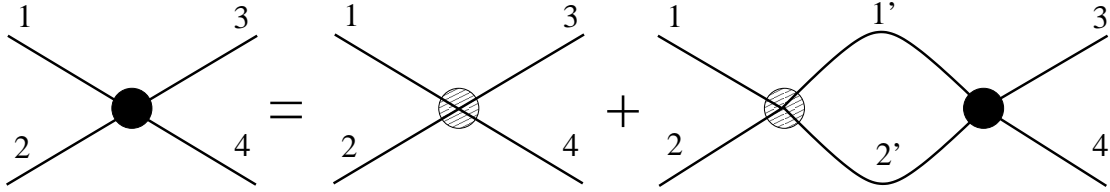}}}
\end{picture}
\caption{The Bethe-Salpeter equation for the two-hole problem. Lines
  represent renormalized Green's functions, dashed circles represent the 
bare ``compact'' vertex function $\Gamma^{fg}$, and 
black circles represent the exact vertex $\hat{\Gamma}^{fg}$.}
\label{BSg_fig}
\end{figure}

If an exact two-particle Green's function has a pole in the scattering
channel at fixed total
energy of the particles, then there exists a bound state
whose energy is the total energy of the particles.
Thus, to analyze the bound state problem of two holes 
one has to solve the Bethe-Salpeter equation, whose most general form
is given by
\begin{eqnarray}
\label{BSg}
&&\hat{\Gamma}^{fg}_{\bf P,k,p}(\Omega,\epsilon,\epsilon^{\prime})=
\Gamma^{fg}_{\bf P,k,p}(\Omega,\epsilon,\epsilon^{\prime})+
\sum_{\bf p_1,\epsilon_1}
\Gamma^{fg}_{\bf P,k,p_1}(\Omega,\epsilon,\epsilon_1)
G_{\bf P/2+p_1}(\Omega/2+\epsilon_1)
G_{\bf P/2-p_1}(\Omega/2-\epsilon_1)
\hat{\Gamma}^{fg}_{\bf P,p_1,p}(\Omega,\epsilon_1,\epsilon^{\prime})\ ,
\end{eqnarray}
where
\begin{eqnarray}
\label{notations}
&&{\bf P}={\bf k_1}+{\bf k_2}={\bf k_3}+{\bf k_4}=
{\bf k_{1^{\prime}}}+{\bf k_{2^{\prime}}}\ , \ \ \
{\bf k}=({\bf k_1}-{\bf k_2})/2\ , \ \ \ 
{\bf p}=({\bf k_3}-{\bf k_4})/2\ , \ \ \ 
{\bf p_1}=({\bf k_{1^{\prime}}}-{\bf k_{2^{\prime}}})/2\ , \nonumber\\
&&\Omega=\epsilon_1+\epsilon_2=\epsilon_3+\epsilon_4
=\epsilon_{1^{\prime}}+\epsilon_{2^{\prime}}\ , \ \ \
\epsilon=(\epsilon_1-\epsilon_2)/2\ , \ \ \
\epsilon^{\prime}=(\epsilon_3-\epsilon_4)/2\ , \ \ \
\epsilon_1=(\epsilon_{1^{\prime}}-\epsilon_{2^{\prime}})/2\ ,
\end{eqnarray}
and the indices are given in accordance with Fig.~\ref{BSg_fig}. 
\noindent
The bare vertex function $\Gamma^{fg}$ includes all diagrams which cannot
be reduced to the second term of Eq. (\ref{BSg}), i.e., those
which cannot be cut by a vertical line between the ends $1$,
$2$ and $3$, $4$ into two parts joined only by two hole lines.
The set of all such diagrams is called a ``compact'' vertex.

At the pole, when $\Omega=E$ ($E$ is the energy of the bound state),
the left-hand side and the second term in the right-hand side of
Eq. (\ref{BSg}) are singular. Therefore $\hat{\Gamma}^{fg}\gg\Gamma^{fg}$, and
the first term in the right-hand side can be neglected. Hence the
integral equation  (\ref{BSg}) becomes homogeneous in 
$\hat{\Gamma}^{fg}$. The variables ${\bf p}$ and $\epsilon^{\prime}$
become parameters and are not
defined by the equation itself. Omitting them and the indices $f, g$
yields 
\begin{eqnarray}
\label{BS1}
\hat{\Gamma}_{\bf P,k}(E,\epsilon)=
\sum_{\bf p_1,\epsilon_1}
\Gamma_{\bf P,k,p_1}(E,\epsilon,\epsilon_1)
G_{\bf P/2+p_1}(E/2+\epsilon_1)
G_{\bf P/2-p_1}(E/2-\epsilon_1)
\hat{\Gamma}_{\bf P,p_1}(E,\epsilon_1)\ .
\end{eqnarray}
Further simplifications of this general formula Eq.~(\ref{BS1})
are possible based on the specifics of the problem being considered. 
Rather general and frequently used simplification can be performed,
for example, when the bare interaction $\Gamma$ is
a potential-like term, i.e. independent of the frequency.
In this case $\hat{\Gamma}$ becomes $\epsilon$-independent and plays
only an auxiliary role. 
Furthermore, it is more convenient to use the bound state
wave function 
$\psi_{\bf P,k}(E)=\sum_{\epsilon}GG\hat{\Gamma}$. By changing the
variables and integrating both sides of
Eq.~(\ref{BS1}) over $\epsilon$ we obtain the Schr\"{o}dinger equation 
for the bound state in momentum representation
\begin{eqnarray}
\label{Sch}
\psi_{\bf P,k}(E)=
\left(\sum_{\epsilon}G_{\bf P/2+k}(E/2+\epsilon)
G_{\bf P/2-k}(E/2-\epsilon)\right)
\sum_{\bf p_1}\Gamma_{\bf P,k,p_1}\psi_{\bf P,p_1}(E)\ .
\end{eqnarray}
If the Green's functions are the bare ones, the sum over $\epsilon$
simply gives $1/[E-\epsilon({\bf P/2+k})-\epsilon({\bf P/2-k})]$,
where $\epsilon({\bf k})$ is the single-particle energy.

In the $t$-$J_z$ model the following characteristics
allow one to simplify Eq.~(\ref{BS1}) considerably and obtain most of
the results for the bound state problem in a transparent analytical
form:\\
(i) The single-hole Green's function is mostly ${\bf k}$-independent,
meaning that in a wide range of $t/J$ ratio the ${\bf k}$-dependent
contribution to the self-energy 
is insignificant or small.\\
(ii) The lowest order contributions to the compact vertex 
$\Gamma_{\bf P,k,p_1}(E,\epsilon,\epsilon_1)$ have a simple kinematic
structure and, together with (i), allow one to separate the momentum and
frequency dependence of the exact vertex $\hat{\Gamma}$.
Its ${\bf k}$-dependent part can be classified
in terms of $s$-, $p$- and $d$-wave harmonics and can be
integrated out easily.\\
(iii) All one-loop and lowest order crossing corrections 
are exactly
zero because of spin conservation in the hole-magnon
interaction. Non-zero corrections are of higher order and 
are expected to be small. Therefore,
the lowest order diagrams in $\Gamma$ can be left 
unrenormalized. \\
(iv) As we have shown in Sec. III the single-hole spectral function
consists of a set of well separated quasiparticle
peaks. Therefore, the Green's function of the hole can always  be
written in the form 
\begin{eqnarray}
\label{G_sum}
G(\epsilon)=\sum_\nu \frac{a_\nu}{\epsilon-
\epsilon_\nu+i0}\ ,
\end{eqnarray}
where $a_\nu$ and $\epsilon_\nu$ are the residue and the energy of the pole
$\nu$, respectively. Then, if the characteristic binding energy is
much smaller than the separation between the quasiparticle peaks
($\Delta=E-2\epsilon_0\ll \delta\epsilon$),
i.e. one half of the bound state energy 
 is close to the lowest peak ($E/2\approx\epsilon_0$), 
one can safely neglect the contribution of the higher poles and use 
the lowest pole approximation for the hole Green's function:
\begin{eqnarray}
\label{G_lp}
G(\epsilon)\approx\frac{a_0}{\epsilon-
\epsilon_0+i0}\ .
\end{eqnarray}
As a matter of fact, in the limiting case $t\gg J$ the separation 
of the levels is $\delta\epsilon\sim J(t/J)^{1/3}$ whereas the binding 
energy, as we will show, is $E\sim -J(J/t)$. Thus,
$E/\delta\epsilon\sim (J/t)^{4/3}\ll 1$ and the lowest pole
approximation should work well. 
In the opposite limit $t\ll J$ the binding energy is $E\sim -J/2$ while
the separation of the quasiparticle poles is $\delta\epsilon\sim
2J$. Besides a factor of the order of
$E/\delta\epsilon\sim 1/4$, the contribution of the higher poles is
strongly reduced because their
quasiparticle residues are negligible ($\sim t^2/J^2$), thus
justifying the lowest pole approximation in this case.  
 One can then argue that the same is
true for all $t/J$. 

With these ideas in mind let us look at the compact
vertex $\Gamma$. 

\begin{figure}
\unitlength 1cm
\epsfxsize=6.5cm
\begin{picture}(8,8)
\put(0,0.7){\rotate[r]{\epsffile{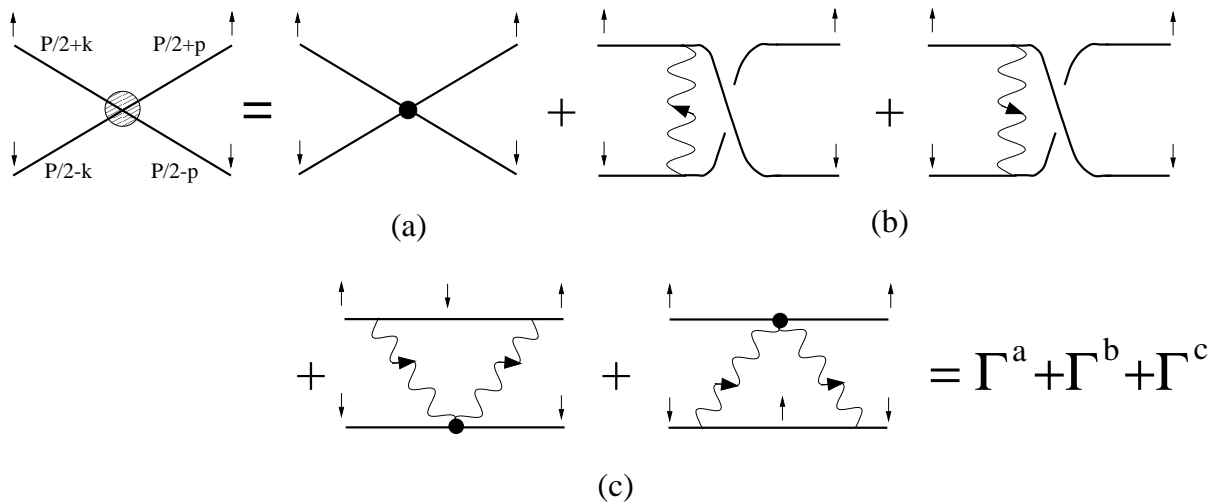}}}
\end{picture}
\caption{The lowest order diagrams contributing to 
the ``compact'' vertex function $\Gamma$: (a) is due to
nearest-neighbor hole-hole attraction, (b) are the one-magnon-exchange
diagrams, (c) are the two-magnon exchange diagrams. (a), (b), and (c)
are of  first, second, and third order in ${\cal H}_{int}$,
respectively. Since the emission of magnon changes the spin of the hole,
the positions of 
the ends of the lines in the diagrams in (b) are interchanged.  
$P/2\pm k=({\bf P}/2\pm{\bf k}, E/2\pm\epsilon)$, $P/2\pm p=({\bf
P}/2\pm{\bf p}, E/2\pm\epsilon_1)$. } 
\label{gam_comp}
\end{figure}
\noindent
The lowest order contributions to the hole-hole compact vertex $\Gamma$
are given in Fig. \ref{gam_comp}.
$\Gamma^a$, $\Gamma^b$, and $\Gamma^c$ are of
first, second, and third order in 
${\cal H}_{int}={\cal H}_1+{\cal H}_2$ (Eq.~(\ref{H1})), respectively. 
They can also be classified 
in terms of powers of $J$ and $t$: $\Gamma^a\sim J$, $\Gamma^b\sim t^2/J$, 
and $\Gamma^c\sim t^2/J$.

\begin{figure}
\unitlength 1cm
\epsfxsize=3.5cm
\begin{picture}(4,4)
\put(0,0.7){\rotate[r]{\epsffile{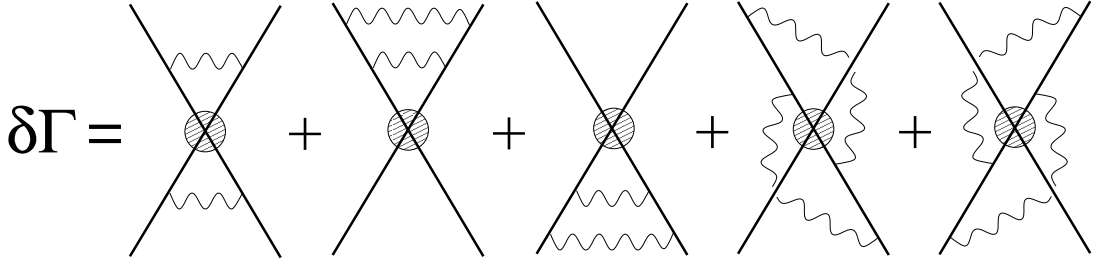}}}
\end{picture}
\caption{Lowest order corrections to 
the ``compact'' vertex function $\Gamma$ (Fig. \ref{gam_comp}).
Circles represent $\Gamma=\Gamma^a+\Gamma^b+\Gamma^c$, wavy lines are
magnons.}
\label{gam_corr}
\end{figure}

If we consider $\Gamma^b$ only,
the full vertex $\hat{\Gamma}$
in Fig. \ref{BSg_fig}\ 
is given by the sum of the ladder diagrams only. To some extent, this 
approximation is analogous to the SCBA in the single-hole problem,
and possesses an important property of the latter: all crossing diagrams and 
one-loop corrections are exactly zero. More importantly, such
corrections  
to {\it all} the lowest order diagrams ($\Gamma^a$, $\Gamma^b$, 
and $\Gamma^c$) are also zero because of the same spin-conservation rule.
For the same reason, higher order corrections to Fig. 
\ref{gam_comp} are strongly suppressed, so that the first dressing of
$\Gamma^a$,  
$\Gamma^b$, and $\Gamma^c$ by magnons appears in the fifth, sixth, and 
seventh order, respectively. 
The corresponding set of diagrams is given in Fig.
\ref{gam_corr}.

\noindent
Note that because all these corrections are of similar origin to the 
Trugman's processes for the self-energy, one can expect that they play only 
a minor role in the effective interaction.

However, there are other sources of correction to $\Gamma$ 
coming from the hole-two-magnon, hole-three-magnon, and
magnon-magnon interactions. Most of them result in the renormalization
of the hole-magnon vertices similar to those in
Fig. \ref{diag3}. While a formal account of such corrections is rather
difficult, one can anticipate the result of such a renormalization
using the real-space consideration of the interactions. As in the
single-hole case, the dressing of the vertex accounts for
two facts. First, the energy of a magnon in the neighborhood of
a hole is $-J/2$ per hole-magnon link lower than the energy of a
free excitation. 
Second, the coordination number for hole hopping 
in a state with a string of magnons is $z-1$
instead of $z$ . 
The first effect
renormalizes the energy of the magnon lines in $\Gamma^b$ and $\Gamma^c$,
while the second effect renormalizes the inner hole line in $\Gamma^c$.

In what follows we will consider the compact vertex $\Gamma$ as
given by $\Gamma^a+\Gamma^b+\Gamma^c$ in Fig. \ref{gam_comp} and will
include the renormalization later.

We first consider the $t=0$ limit when the solution of the BS
equation (\ref{BS1}) is trivial. In this limit only
$\Gamma^a=-2J\gamma_{\bf k_1-k_3}$ survives and the hole Green's
function is simply $1/\omega$. Thus Eq.~(\ref{BS1}) becomes:
\begin{eqnarray}
\label{t_0}
\hat{\Gamma}_{\bf P,k}(E,\epsilon)=-2J
\sum_{\bf p_1}\gamma_{\bf k-p_1}
\sum_{\bf \epsilon_1}
\frac{\hat{\Gamma}_{\bf P,p_1}(E,\epsilon_1)}
{(E/2+\epsilon_1+i0)(E/2-\epsilon_1+i0)}\ .
\end{eqnarray}
Evidently, the dependence of $\hat{\Gamma}$ on ${\bf P}$ is arbitrary and
$\hat{\Gamma}(\epsilon)$ is a constant. Integrating over $\epsilon_1$ one
finds:
\begin{eqnarray}
\label{t_0_a}
\hat{\Gamma}_{\bf k}(E)=-\frac{2J}{E}
\sum_{\bf p_1}\gamma_{\bf k-p_1}
\hat{\Gamma}_{\bf p_1}(E)\ ,
\end{eqnarray}
where $\hat{\Gamma}$ plays the role of the wave function.
Since the ${\bf k}$-structure of the vertex is very simple, 
one can classify all solutions of Eq.~(\ref{t_0_a}) in terms of the
nearest-neighbor $s$, $p$, and $d$ waves: 
$\varphi_{\bf k}^s=\cos(k_x)+\cos(k_y)$, $\varphi_{\bf k}^p=\sqrt{2}
\sin(k_x);\sqrt{2}\sin(k_y)$, $\varphi_{\bf
k}^d=\cos(k_x)-\cos(k_y)$. Using the property: $\sum_{\bf p}
\gamma_{\bf k-p}\varphi_{\bf p}^{s(p,d)}=\frac{1}{4}\varphi_{\bf k}^{s(p,d)}$,
one can see that these solutions are degenerate with
energy $\Delta^{s,p,d}=E^{s,p,d}=-J/2$. 
This is expected if one remembers that while an
individual hole breaks four AF bonds, 
two nearest-neighbor holes save one bond.

A finite hopping constant
$t$ has three effects: (i) the hole Green's function becomes
renormalized, (ii) $\Gamma^b$ and $\Gamma^c$ start to contribute to
the interaction, (iii) the hole self-energy also acquire a small ${\bf
k}$-dependent contribution.
In order to see their roles,
let us incorporate these changes into the BS equation (\ref{BS1}) one by
one. 

\begin{figure}
\unitlength 1cm
\epsfxsize=9cm
\begin{picture}(9,8.5)
\put(0,-3){\epsffile{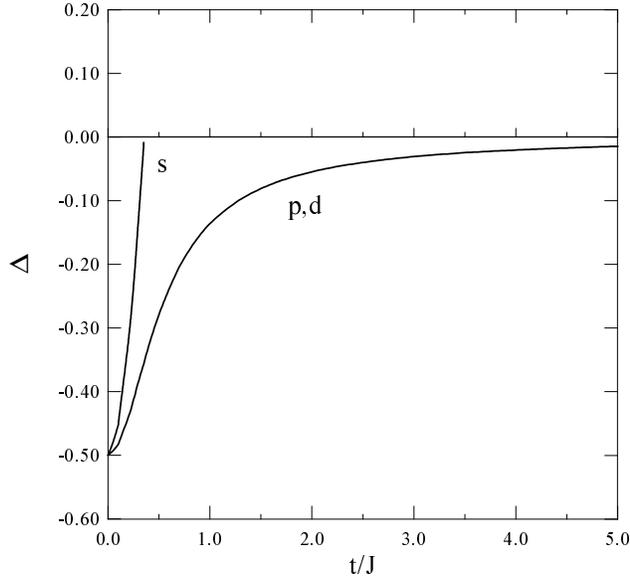}}
\end{picture}
\caption{Binding energies (in units of $J$)
of the $s$, $p$, and $d$ bound states.
When only the contact interaction $\Gamma^a$ is taken into account,
all three bound states are degenerate (Eq.~(\ref{t_1a})) and are shown by the
lower curve.
When the contact and magnon-exchange
interactions 
$\Gamma^a+\Gamma^b$ are considered, the $p$ and $d$ energies are
unchanged (lower curve) while the energy of the $s$-state is pushed up 
(Eq.~(\ref{t_2_e})) (upper curve). 
}
\label{loc_is1}
\end{figure}

Consider the case when the hole Green's function is renormalized as
described in Sec. III and is ${\bf k}$-independent, but the interaction 
still comes from the
$\Gamma^a$ term only. Since $\Gamma^a$ is ${\bf P}$ and $\epsilon$
independent, the BS equation is similar to 
Eq.~(\ref{t_0_a})
\begin{eqnarray}
\label{t_1}
\hat{\Gamma}_{\bf k}(E)=-2J
\sum_{\bf p_1}\gamma_{\bf k-p_1}
\hat{\Gamma}_{\bf p_1}(E)\sum_{\epsilon_1}
G(E/2+\epsilon_1)G(E/2-\epsilon_1)\ .
\end{eqnarray}
Since the ${\bf k}$-structure of the core is unchanged,
all three harmonics remain degenerate. Hence
\begin{eqnarray}
\label{t_1_a}
1=-\frac{J}{2}\sum_{\epsilon_1}G(E/2+\epsilon_1)G(E/2-\epsilon_1)\ .
\end{eqnarray}
In the lowest-pole approximation $G$ is given by Eq.~(\ref{G_lp})
so the binding energy relative to the bottom of the hole band is:
\begin{eqnarray}
\label{t_1a}
\Delta=E-2\epsilon_0=-J a_0^2/2\ . 
\end{eqnarray}
In the limit $t\ll J$,
$\Delta\approx -J(1-16t^2/9J^2)/2$. In the limit $t\gg J$,
 $a_0\sim J/t$ so $\Delta\sim -J(J/t)^2$. The binding energy (in units of $J$)
v.s. $t/J$ is shown in Fig.~\ref{loc_is1}. This energy is always
negative because we assume that the hole forms
a completely localized excitation. In this case a bound state is always
formed as long as the effective interaction remains attractive.

Now we consider the magnon exchange interaction
$\Gamma^b$ which is given by:
\begin{eqnarray}
\label{t_2}
\Gamma^b_{\bf k,p,P}(\epsilon,\epsilon_1)=-16t^2\bigg[
\frac{\gamma_{\bf P/2-p}\gamma_{\bf P/2-k}}{\epsilon+\epsilon_1-2J+i0}+
\frac{\gamma_{\bf P/2+p}\gamma_{\bf P/2+k}}{-\epsilon-\epsilon_1-2J+i0}
\bigg]\ .
\end{eqnarray}
Since $\Gamma^b$ depends on the total momentum of the pair ${\bf P}$,
the solutions of the BS equation also depends on  ${\bf P}$. The energy of
the bound state at different ${\bf P}$ provides a dispersion law
for the pair.
In general, the lowest energy bound state is always at
zero or some finite momenta corresponding to high symmetry points 
in the ${\bf k}$-space. In
this work we do not consider finite-${\bf P}$ bound states because
their physics is unimportant for the purposes of this paper.

The BS equation (\ref{BS1}) at ${\bf P}=0$ with interaction
$\Gamma^a+\Gamma^b$ reads:
\begin{eqnarray}
\label{t_2_a}
&&\hat{\Gamma}_{\bf 0,k}(E,\epsilon)=-2J
\sum_{\bf \epsilon_1}G(E/2+\epsilon_1)G(E/2-\epsilon_1)\\
&&\phantom{\hat{\Gamma}_{\bf 0,k}(E,\epsilon)=-2J}
\times
\sum_{\bf p}\bigg\{\gamma_{\bf k-p}+\frac{8t^2}{J}
\gamma_{\bf p}\gamma_{\bf k}\bigg[
\frac{1}{\epsilon+\epsilon_1-2J+i0}+
\frac{1}{-\epsilon-\epsilon_1-2J+i0}
\bigg]\bigg\}\hat{\Gamma}_{\bf 0,p}(E,\epsilon_1)\ .
\nonumber
\end{eqnarray}
Evidently, the ${\bf k}$ and $\epsilon$ dependencies of $\hat{\Gamma}$
can be separated and one can see that the $p$- and $d$-wave
solutions are not affected by the magnon-exchange interaction
because $\sum_{\bf p}\gamma_{\bf p}\varphi_{\bf p}^{p(d)}\equiv 0$. This
observation can be expressed in a different form: in the zero total
momentum 
$p$- or $d$-wave state the amplitudes of the spin-flip exchange from
different lobes of the wave function cancel each other.

For the $s$-wave state, substituting $\hat{\Gamma}_{\bf
k}(\epsilon)=\varphi_{\bf k}^s\hat{\Gamma}(\epsilon)$ and
integrating over ${\bf k}$ give:
\begin{eqnarray}
\label{t_2_b}
\hat{\Gamma}_E(\epsilon)=-\frac{J}{2}
\sum_{\bf \epsilon_1}G(E/2+\epsilon_1)G(E/2-\epsilon_1)
\bigg\{1+\frac{8t^2}{J}\bigg[
\frac{1}{\epsilon+\epsilon_1-2J+i0}+
\frac{1}{-\epsilon-\epsilon_1-2J+i0}
\bigg]\bigg\}\hat{\Gamma}_E(\epsilon_1)\ .
\end{eqnarray}
In the so called static approximation, when the retardation due to the
magnon propagator is neglected, $\Gamma^b$ is
$\epsilon$-independent and Eq.~(\ref{t_2_b}) becomes
\begin{eqnarray}
\label{t_2_c}
1=-\frac{J}{2}\bigg\{1-\frac{8t^2}{J}\bigg\}\sum_{\bf \epsilon_1}
G(E/2+\epsilon_1)G(E/2-\epsilon_1)\simeq
-\frac{J}{2}\frac{a_0^2}{\Delta^s}\bigg\{1-\frac{8t^2}{J^2}\bigg\} \ .
\end{eqnarray}
As in the previous case,
 $\Delta^s=E^s-2\epsilon_0$. The problem with retardation can
be solved using a method introduced in Ref.~\cite{BCS}. One can assume
that $\hat{\Gamma}_E(\epsilon)$ has no singularities
 \cite{comm1} and therefore the integral over
$\epsilon_1$ in Eq.~(\ref{t_2_b}) is defined by the poles of $G$ and the
magnon propagator only. In the lowest-pole approximation, by using the
obvious parity $\hat{\Gamma}(-\epsilon)=\hat{\Gamma}(\epsilon)$ and by
choosing the proper contour of integration one obtains:
\begin{eqnarray}
\label{t_2_d}
\hat{\Gamma}_E(\epsilon)=-\frac{J}{2}\frac{a_0^2}{\Delta^s}
\bigg\{1+\frac{8t^2}{J}\bigg[
\frac{1}{\epsilon+\Delta^s/2-2J+i0}+
\frac{1}{-\epsilon-\Delta^s/2-2J+i0}
\bigg]\bigg\}\hat{\Gamma}_E(\Delta^s/2)\ .
\end{eqnarray}
Multiplying both sides by $G(E/2+\epsilon)G(E/2-\epsilon)$ and
integrating over $\epsilon$ finally yields:
\begin{eqnarray}
\label{t_2_e}
1=-\frac{J}{2}\frac{a_0^2}{\Delta^s}\bigg\{1+\frac{16t^2}{J(\Delta^s-2J)}
\bigg\} \ .
\end{eqnarray}
This expression coincides with that given by the Wigner-Brillouin
perturbation theory. The resulting
binding energies (from  $\Gamma^a+\Gamma^b$)
v.s. $t/J$ are shown in Fig. \ref{loc_is1}.

\noindent
Thus the $s$-wave binding energy vanishes at
$(t/J)|_c=1/(2\sqrt{2})\simeq 0.35$. Since the spin-polarons 
are assumed to be localized the disappearance of the bound
state is not due to the competition between the delocalizing kinetic energy
and an attraction, but rather due to the competition between different
types of interaction. One can see that the magnon exchange actually
leads to an effective repulsion between the holes. In a naive
point of view the holes can be mutually attracted
by following each other in a
caterpillar-like process, because in this way the disturbance on the AF
background is only virtual. However, this scenario does not take into
account the fact that after an elementary hopping of a hole pair,
the order of the holes is different from the initial one, 
i.e., they are exchanged \cite{comm2}. 
This in turn makes
the corresponding energy correction positive, i.e. it leads to repulsion.

It is interesting to note that while the hole-magnon interaction
plays a major role in the hole dressing, it has no direct effect on the
$p$ and $d$ bound states and is destructive for the $s$-wave state.

\begin{figure}
\unitlength 1cm
\epsfxsize=1cm
\begin{picture}(2,2)
\put(4,0.7){\rotate[r]{\epsffile{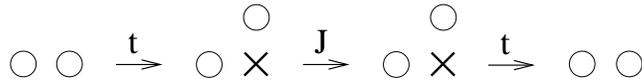}}}
\end{picture}
\caption{Real-space image of the process in Fig. \ref{gam_comp}(c)}
\label{real_space}
\end{figure}

Let us now add the third part of the hole-hole interaction
$\Gamma^c$ and see if the combination of the hole-magnon interaction
with the two-magnon vertex causes an additional attraction.
The diagrams in Fig. \ref{gam_comp}(c) are equivalent to:
\begin{eqnarray}
\label{t_3}
\Gamma^c_{\bf k,p}(E,\epsilon,\epsilon_1)=-2J (4t)^2\gamma_{\bf k-p}
\sum_{\bf q}\gamma_{\bf k-q}^2\sum_{\omega}\frac{1}{\omega-2J+i0}
\bigg[
\frac{G(\epsilon+E/2-\omega)}{\omega-\epsilon+\epsilon_1-2J+i0}+
\frac{G(-\epsilon+E/2-\omega)}{\omega+\epsilon-\epsilon_1-2J+i0}
\bigg]\ .
\end{eqnarray}
Integration over the inner momentum ${\bf q}$ and frequency $\omega$
yields:
\begin{eqnarray}
\label{t_3_1}
\Gamma^c_{\bf k,p}(E,\epsilon,\epsilon_1)=-8Jt^2\gamma_{\bf k-p}
\frac{1}{\epsilon_1-\epsilon}
\bigg[
G(\epsilon+E/2-2J)-G(-\epsilon+E/2-2J)-(\epsilon\rightarrow\epsilon_1)
\bigg]\ .
\end{eqnarray}
One can see that the kinematic structure of this vertex is identical
to $\Gamma^a$, i.e. this term is effectively a contact
nearest-neighbor interaction. This is evident from the real-space
picture of the process, which is shown in
Fig. \ref{real_space}.

\noindent
Initially the holes are at nearest-neighbor sites. One of
them jumps to a neighboring site creating a spin-flip at its site of
origin. Then the second hole virtually absorbs and reemits
this magnon. Finally the first hole jumps back and the original state
is restored. Naturally, such an interaction depends only on
$\gamma_{\bf k-p}$ and does not depend on ${\bf P}$. One can also
guess the character of the resulting interaction. 
Being in the neighborhood of another hole, the spin excitation
generated by the hopping of the first hole lowers its energy. 
Therefore the interaction between the holes
via such a process must also
be attractive.  

In the lowest-pole approximation $\Gamma^c$ Eq.~(\ref{t_3_1}) is given by:
\begin{eqnarray}
\label{t_3_2}
\Gamma^c_{\bf k,p}(E,\epsilon,\epsilon_1)=-8Jt^2 a_0\gamma_{\bf k-p}
\bigg[\frac{1}{(\epsilon+\Delta/2-2J+i0)(\epsilon_1+\Delta/2-2J+i0)}+
(\epsilon,\epsilon_1\rightarrow-\epsilon,-\epsilon_1)
\bigg]\ .
\end{eqnarray}
One can consider the static limit of the problem by eliminating the
frequency dependence of the propagators in Eq.~(\ref{t_3_2}). This gives
\begin{eqnarray}
\label{t_3_3}
\Gamma^c_{\bf k,p}\simeq-4\frac{t^2}{J} a_0\gamma_{\bf k-p}
\ .
\end{eqnarray}
In the $t\gg J$ limit this expression yields $\Gamma^c_{\bf
k,p}\simeq-4t\gamma_{\bf k-p}=-(2t/J)\Gamma^a_{\bf k,p}\gg
\Gamma^a_{\bf k,p}$. Thus, in this limit the effective interaction
constant of the contact interaction is renormalized from $J$ to
$t$. For the $p$- and $d$-waves it gives a bound state energy
$\Delta^{p(d)}\sim -t a_0^2\sim -J(J/t)$, which is a factor of 
$(t/J)\gg 1$ larger than the result obtained with $\Gamma^a$ only. 
This result, while being quite simple, is somewhat unexpected and, to
our knowledge, has never been discussed in the literature.

A rigorous account for the retardation yields modified equations for
the bound state energies, leaving the qualitative picture drawn in the
static limit essentially unchanged:
\begin{eqnarray}
\label{t_3_4}
&&1=-\frac{J}{2}\frac{a_0^2}{\Delta^{p(d)}}\bigg\{1+\frac{8t^2a_0}
{(\Delta^{p(d)}-2J)^2}\bigg\}\hskip  1cm p{\rm -}, d{\rm -wave},\\
&&1=-\frac{J}{2}\frac{a_0^2}{\Delta^s}\bigg\{1+\frac{8t^2a_0}
{(\Delta^s-2J)^2}+\frac{16t^2}{J(\Delta^s-2J)}\bigg\}
\hskip 1cm s{\rm -wave}
\ .
\nonumber
\end{eqnarray}
Thus the energies of the $p$- and $d$-wave states are always
negative. In the $s$-wave bound state,
$\Gamma^c$ cannot overcome the magnon-exchange repulsion
$\Gamma^b$. It also has  little effect
on $(t/J)|_c^s$ because in the small $t$ region where
$\Delta^s$ goes to zero, $\Gamma^c$ is roughly 4 times smaller than
$\Gamma^b$. 

We now consider the effect of the renormalization. 
As we noted before we have enough reasons to
rule out the possibility that the Trugman-like diagrams Fig. \ref{gam_corr} 
can change the interaction constants significantly. 
The anticipated smallness of these diagrams in the $t>J$ region is
due to the geometrical factor, which is of the order of $1/z^3=1/64$. Most of
these corrections are relatively easy to evaluate using the same
approximations as above. The results of these calculations show
that those terms with the same kinematic structure as in $\Gamma^a$,
$\Gamma^b$, and $\Gamma^c$ are indeed suppressed by this small
factor. They also show that among the corrections there are
terms with a more complicated kinematic structure than the
original $\Gamma$. Since these terms are not suppressed by the
small
factor, they might enhance the effective interaction. Some of these
terms correspond to the contact interaction between the holes in
states where both of them have ``tails'' of strings. Evidently, this
leads to interaction of polarons at sites farther than
nearest neighbors. For the bound state wave function it would mean that
harmonics higher than the nearest-neighbor one should be
considered. Nevertheless, we do not expect 
these terms to make any qualitative change
and in the following we will omit all
corrections from Fig. \ref{gam_corr} to avoid complications. 

\begin{figure}
\unitlength 1cm
\epsfxsize=9cm
\begin{picture}(9,8.5)
\put(0,-3){\epsffile{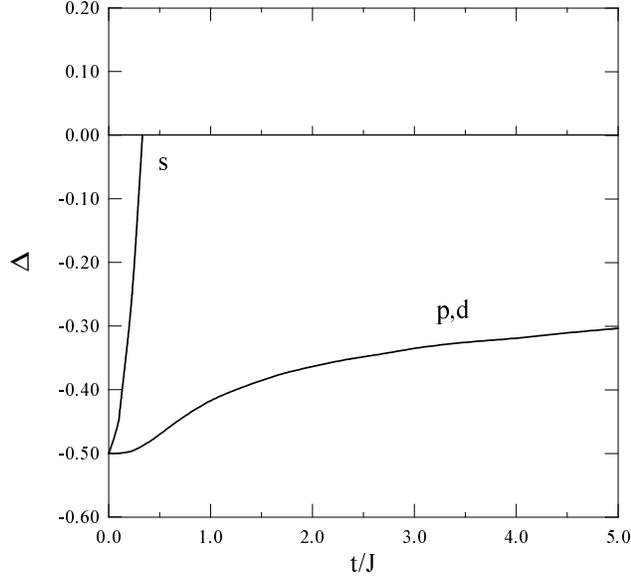}}
\end{picture}
\caption{Binding energies of the $s$, $p$, and $d$ bound states when
all three interactions shown in Fig. \ref{gam_comp} are 
taken into account and the energy of the magnons in the intermediate state
is renormalized.
}
\label{loc_is2}
\end{figure}

At the same time, diagrams of the non-SCBA type
(Fig. \ref{diag3}) are straightforward to include using the real-space
language. We have to take into account the 
lower energy of the magnons in the intermediate state in the diagrams
Fig. \ref{gam_comp}(b,c). This simply changes $\omega_0=2J$ in the
denominators of Eqs.~(\ref{t_3_4}) to $\tilde{\omega}_0=J$. 
Renormalization of the hole energy in the intermediate state in
$\Gamma^c$ produces much weaker effect. This renormalization finally
yields: 
\begin{eqnarray}
\label{t_3_5}
&&\frac{\Delta^{p(d)}}{J}=-\frac{a_0^2}{2}\bigg\{1+\frac{8t^2a_0}
{(\Delta^{p(d)}-J)^2}\bigg\},\\
&&\frac{\Delta^s}{J}=-\frac{a_0^2}{2}\bigg\{1+\frac{8t^2a_0}
{(\Delta^s-J)^2}+\frac{16t^2}{J(\Delta^s-J)}\bigg\}
\ .
\nonumber
\end{eqnarray}
Solutions of these equations are plotted in Fig. \ref{loc_is2}. One
can see that the energies of the $p$ and $d$ bound 
states are changed significantly from those in Fig. \ref{loc_is1}. 

Thus, the hierarchy of
terms in the effective hole-hole interaction emerged from our study
is as follows: 
\\
\noindent
(i) The contact ``sharing common link'' attraction
yields bound states with $s$-, $p$-, $d$-symmetry with energy
$\sim -J/2$ at small $t/J$;\\
\noindent
(ii) The magnon-exchange interaction has no effect on 
the ${\bf P}=0$\ \ $p$- and
$d$-wave states because of symmetry reasons, and is strongly repulsive
for the $s$-wave state;
\\
\noindent
(iii) The hole-magnon attraction, which leads to an attraction between a
hole and a string of another hole, is the strongest interaction in the present
problem in the physical limit ($t> J$). It leads to strong $p$
and $d$ bound states with energy $\sim -J(J/t)$.

Now, after all essential interactions are taken into account and
carefully analyzed, the last effect which has to be incorporated into
the BS equation before we can make comparison with numerical results
is the ${\bf k}$-dependent contributions to the hole Green's function.
As noted in Sec. III such contributions are small, but lead to
a coherent hole band. Therefore the ED data show some dispersion in the
hole energy. For instance, at $t/J=5$ the energy difference between the
${\bf k}=(0,0)$ and ${\bf k}=(\pi/2,\pi/2)$ points (almost exactly
half of the bandwidth) is $W/2\simeq 0.078t=0.39J$. Although this difference is
small compared to the absolute value of the lowest pole
energy, $\epsilon_0\simeq-2.8t$, or the separation between the energy levels,
$\delta\epsilon\sim t$,
it is as large as the hole-hole binding energy calculated above,
$\Delta\sim -0.3J$. Therefore, one would expect this dispersion
to have strong influence
on the pairing of holes. One can also foresee that such an
influence is likely to be destructive, because now the pairing is not
only an interplay between repulsions and attractions,
but also the question of how much kinetic energy is lost if the holes are
in a bound state.

\begin{figure}
\unitlength 1cm
\epsfxsize=9cm
\begin{picture}(9,8.5)
\put(0,-4){\epsffile{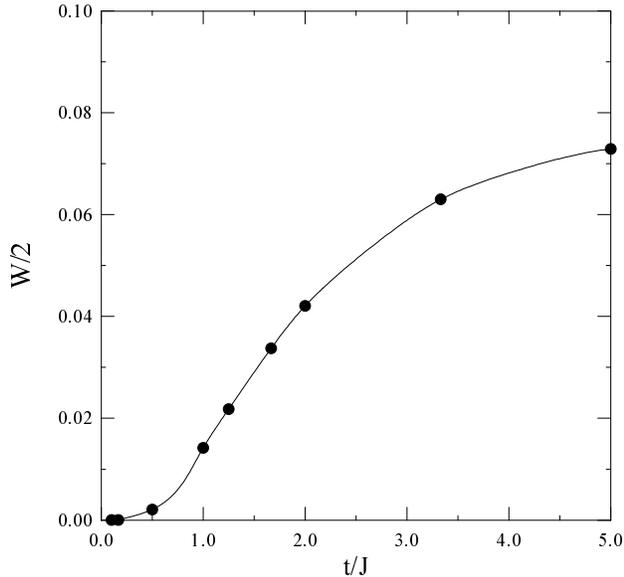}}
\end{picture}
\caption{Half of the hole bandwidth in units of $t$ as the function of
$t/J$. 
}
\label{W}
\end{figure}

For our purposes we will need the hole Green's function in the
lowest-pole approximation, which now reads:
\begin{eqnarray}
\label{G_lp_k}
G(\epsilon)\approx\frac{a_0}{\epsilon-
\tilde{\epsilon_0}-W\epsilon_{\bf k}/2+i0}\ ,
\end{eqnarray}
where $\tilde{\epsilon_0}=\epsilon_0-W/2$, $\epsilon_{\bf
k}=(1-\cos(k_x)\cos(k_y))>0$, $W>0$, and we have neglected the ${\bf
k}$-dependence of $a_0$. The role of the Trugman's loops larger than an
elementary square is negligible, so we consider the dispersion given
by the next-nearest-neighbor hopping only. The resulting band is isotropic
and has a minimum at the center of the BZ.
Now in the BS equation (\ref{BS1}) it is easier to integrate over
$\epsilon_1$ first. Neglecting the effect of the dispersion on the
retardation and after some algebra one obtains:
\begin{eqnarray}
\label{t_4}
\varphi_{\bf k}=-2Ja_0^2\sum_{\bf p}
\frac{\varphi_{\bf p}}{\Delta-W\epsilon_{\bf p}}
\bigg\{\gamma_{\bf k-p}
\bigg[1+\frac{8t^2a_0}{(\Delta-J)^2}\bigg]
+\gamma_{\bf k}\gamma_{\bf p}\frac{16t^2}{J(\Delta-J)}\bigg\}
\ .
\end{eqnarray}
where $\varphi_{\bf k}$ is a nearest-neighbor $s$-, $p$-, or $d$-wave. As
before the magnon-exchange interaction (the second term in the curly
bracket) is orthogonal to the $p$- and
$d$-wave states.
Eq.~(\ref{t_4}) is an almost evident generalization of
Eqs.~(\ref{t_3_5}) to the case of finite hole dispersion. The
only new parameter introduced is the hole bandwidth $W$. Our
numerical ED data for $W/2t$ as a function of $t/J$ 
calculated on a 32-site lattice are given in
Fig. \ref{W}.
We do not calculate this quantity analytically. Instead
we treat it as a parameter and use the numerical values in
Fig \ref{W}.

Two statements can be made about the solutions of  Eq.~(\ref{t_4}).
Now all three symmetries have thresholds $(t/J)|_c^{s,p,d}$ 
above which negative
energy solution of the corresponding symmetry cease to exist.
Also, the $p$- and $d$-waves are split with the $p$ wave state having
lower energy.  

The first statement
is evident if one consider the threshold condition ($\Delta=0$)
for the $p$- or $d$-wave state in Eq.~(\ref{t_4}), assuming $t\gg J$ so
that $a_0\simeq J/t$:
\begin{eqnarray}
\label{t_4_1}
1\simeq -\frac{J}{2}\frac{J^2}{t^2}\sum_{\bf p}
\frac{\varphi_{\bf p}^{p,d}}{-W\epsilon_{\bf p}}
\bigg\{1+\frac{8t^2}{J^2}\frac{J}{t}\bigg\}\simeq
\frac{4J}{W}\frac{J}{t}
\sum_{\bf p}\frac{\varphi_{\bf p}^{p,d}}{(1-\cos(k_x)\cos(k_y))}
\ .
\end{eqnarray}
The sum over ${\bf p}$ gives a number of the order of unity. Thus, there
is a value of $W$ ($(W/t)_c\simeq 4J^2/t^2\ll 1$) at which the
equality condition is satisfied. If $W$ grows with $t$, as it in fact
does, then above this point
Eq.~(\ref{t_4}) has no solution. 
Using  the dependence of $W$ on $t/J$ (Fig. \ref{W}),
the critical values of $t/J|_c^{p,d}$ can be extracted from the same
condition Eq.~(\ref{t_4_1}).

The second statement is a consequence of the isotropic form of the
dispersion law. The sum over ${\bf p}$ in Eq.~(\ref{t_4_1}) can be
rewritten as: 
\begin{eqnarray}
\label{t_4_2}
A^{p,d}=\sum_{\bf p}
\frac{\varphi_{\bf p}^{p,d}}{\epsilon_{\bf p}}=\sum_{\bf p}
\varphi_{\bf p}^{p,d}[1+\cos(k_x)\cos(k_y)+(\cos(k_x)\cos(k_y))^2
+\cdots ]=1+C_1+C_2+\cdots
\end{eqnarray}
In the case of $d$ wave the series is alternating with $C_1=-1/2$ and
$C_2=3/8$. In the case of $p$ wave $C_n\equiv 0$ for odd $n$, and
$C_n>0$ for even $n$. Numerical integration gives $A^p=1.27$,
$A^d=0.73$, and therefore 
$d$ wave solution disappears at a lower value of $t/J$.

\begin{figure}
\unitlength 1cm
\epsfxsize=9cm
\begin{picture}(9,8.5)
\put(0,-3){\epsffile{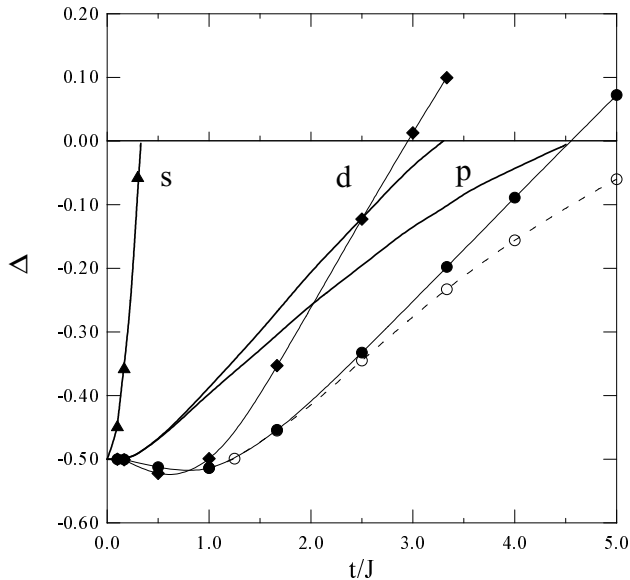}}
\end{picture}
\caption{Binding energies of the $s$-, $p$-, and $d$-wave bound states. Bold
lines are the solutions of Eq.~(\ref{t_4}). Triangles, diamonds, and
solid circles are the 32-site ED results for the $s$-, $d$- and
$p$-wave, respectively. Lines connecting these points are
guides for the eye. Empty circles are the results of modified Lanczos
study on a 50-site cluster from Ref.~\protect\cite{RD}.}
\label{s_p_d}
\end{figure}

Numerical solutions of Eq.~(\ref{t_4}) for all three symmetries
together with results of our 32-site ED calculations are shown in
Fig. \ref{s_p_d}. For the purpose of comparison we also show
the results obtained by a modified Lanczos method on a 50-site cluster
in Ref. \cite{RD}.

\noindent
One can see a very good {\it quantitative} agreement between our
analytical and numerical results for the $s$-wave state.
For the $d$- and $p$-states, they have the same qualitative behaviors. 
It is clear that our Eq.~(\ref{t_4}), while involving different kinds of
approximations, captures all essential effects of the pairing problem
in the $t$-$J_z$ model in a very good quantitative level. It also
describes the problem in a very easy and transparent way. 
Namely, all essential tendencies in the hole binding are described by
this equation: (i) the $s$-wave is pushed up to the continuum by the
magnon-exchange process, (ii) the $p$- and $d$-waves are much more stable,
(iii) the Trugman's dispersion split these two levels and push them
to the continuum at some larger critical values of $t/J$.

The analytical results underestimate the pairing strength,
especially in the region $t/J\sim 1$. We believe that this is
due to the 
contribution from the higher poles which are neglected in our study. 
In this region a significant portion of the spectral weight is
transfered to these poles while the separation between them 
is still of the order of $2J$. Since our lowest-pole approximation uses the
ratio $\Delta/\delta\epsilon$ ($\simeq 1/4$ in this region) 
as the small parameter, one would expect the energy correction from
the higher poles to be of the same order.
In the region where the binding energy is close to zero we expect
this approximation to be more accurate. However, for the $p$ and
$d$ waves this is also the region of $t/J>2$ where the omitted higher order
diagrams with more complicated kinematic structure contribute to the
interaction. We refer the smaller splitting of the $p$- and $d$-wave
states found in our analytical results compared to those in our ED data to the
effect of these corrections. The same higher order interactions
cause a tiny splitting of the states at $t/J<1$ which is shown in
Fig. \ref{s_p_d}. We remark that a $t/J$ dependence of the binding
energies very similar to our ED results has been found recently by
series-expansion calculations \cite{Hamer}. 
Thus we believe that the agreement between the results can be
improved by an accurate consideration  
of the contribution of higher poles and higher order diagrams
of various types. However, this is out of the scope of our paper.

It is natural and instructive to consider the case where
small transverse spin fluctuations $J_{\perp}=\alpha J$ (with $\alpha\ll 1$)
are included in the Hamiltonian. 
This helps one to see the effects of these
fluctuations on the pairing problem and to infer the tendencies in
the isotropic $t$-$J$ model. As noted in Ref. \cite{Star} only
two changes are necessary to account for these fluctuations
up to first order in $\alpha$. First, the holes acquire 
a dispersion law given by
\begin{eqnarray}
\label{t_5}
\delta\epsilon_{\bf k}=\alpha A\gamma_{\bf k}^2
\end{eqnarray}
where $A$ is of the order of $t^2/J$ in the $t\ll J$ limit and of the order
of $J$ in the $t\gg J$ limit \cite{Star}. Such a band is strongly
anisotropic, i.e. 
degenerate minima form a line along the magnetic BZ boundary.
In the full $t$-$J$ model ($\alpha=1$) this dispersion
dominates over the Trugman's terms at all $t/J$.
Second, the hole-magnon interaction becomes:
\begin{eqnarray}
\label{t_5_1}
\Gamma_{\bf k,q}^{(1)}=4t[\gamma_{\bf k-q}-\frac{1}{2}\alpha
\gamma_{\bf k}\gamma_{\bf q}+O(\alpha^2)] \ .
\end{eqnarray}
Corrections to the magnon energy are of higher order in $\alpha$:
$\omega_0=2J+O(\alpha^2)$. As we shall see both changes lead to the
splitting of the $p$- and $d$-wave levels, and both favor the
$d$-wave as the ground state of the system. 

Consider Eq.~(\ref{t_4}) with the dispersion from Eq.~(\ref{t_5}) and
the hole-magnon exchange term coming from Eq.~(\ref{t_5_1}). For the
purpose of illustration it is sufficient 
to consider the interactions in the static
limit:
\begin{eqnarray}
\label{t_5_2}
\varphi_{\bf k}=-2Ja_0^2\sum_{\bf p}
\frac{\varphi_{\bf p}}{\Delta-2\alpha A\gamma_{\bf p}^2}
\bigg\{\gamma_{\bf k-p}
\bigg[1+\frac{8t^2a_0}{J^2}\bigg]
-\frac{16t^2}{J^2}\bigg[\gamma_{\bf k}\gamma_{\bf p}-\frac{\alpha}{2}
\gamma_{\bf k+p}(\gamma_{\bf k}^2+\gamma_{\bf p}^2)\bigg]
\bigg\}
\ .
\end{eqnarray}
Corrections to the bound state energies of the $p$ and
$d$ wave up to first order in $\alpha$ are:
\begin{eqnarray}
\label{t_5_3}
&&\Delta^d=\Delta_0^d+2\alpha\bigg(A-\frac{2t^2}{J}a_0^2\bigg)
\sum_{\bf p}(\varphi_{\bf p}^d)^2\gamma_{\bf p}^2=
\Delta_0^d+\frac{\alpha}{8}\bigg(A-\frac{2t^2}{J}a_0^2\bigg)\ , \\
&&\Delta^p=\Delta_0^p+2\alpha\bigg(A+\frac{2t^2}{J}a_0^2\bigg)
\sum_{\bf p}(\varphi_{\bf p}^p)^2\gamma_{\bf p}^2=
\Delta_0^p+\frac{3\alpha}{8}\bigg(A+\frac{2t^2}{J}a_0^2\bigg)\ ,
\nonumber
\end{eqnarray}
where the first term in the bracket comes from the dispersion and the
second is from the magnon-exchange interaction. The kinetic energy pushes up
the energy of both states, but for the $p$-wave this shift is three
times larger. An important observation is that the magnon-exchange
interaction is now attractive
for the $d$-wave and repulsive for the $p$-wave state!

To show how subtle the selection between the $p$- and
$d$-wave ground state is, let us consider a more "realistic" dispersion
law: 
\begin{eqnarray}
\label{t_5_4}
\delta\epsilon_{\bf k}=\frac{\alpha}{4} A(\cos(k_x)^2+\cos(k_y)^2)\ .
\end{eqnarray}
It is more realistic in the sense that it has 
minima at $\pm(\pm\pi/2,\pi/2)$ points which are isotropic and it
agrees with the experimentally measured profile in the cuprates
better than the one in Eq.(\ref{t_5}). 
It is now known that this 
difference between the
$t$-$J$ model dispersion law (Eq.(\ref{t_5})) and the one observed in
reality (Eq.(\ref{t_5_4}))
is due to additional intrasublattice hoppings
($t^\prime, t^{\prime\prime}$ etc.) in the real systems \cite{Naz}.  
With the dispersion law in Eq.~(\ref{t_5_4}) corrections to the binding
energies are given by: 
\begin{eqnarray}
\label{t_5_5}
&&\Delta^d=\Delta_0^d+\frac{\alpha}{8}\bigg(5A-\frac{2t^2}{J}a_0^2\bigg)\ , \\
&&\Delta^p=\Delta_0^p+\frac{\alpha}{8}\bigg(3A+\frac{2t^2}{J}a_0^2\bigg)\ .
\nonumber
\end{eqnarray}
Now the kinetic energy pushes up the $d$-wave energy $5/3$ times more than
that of the $p$-wave and $\Delta^p-\Delta^d=\frac{\alpha}{4}[2t^2a_0^2/J-A]$
is not necessarily positive. Therefore, $p$-wave can become a ground state
again. 

Summarizing our observations in the small $\alpha$ limit,
we conclude that the transverse fluctuations create an
additional attractive interaction in the $d$-symmetry  state and
that an anisotropic kinetic energy profile also favors $d$-wave. Thus, we
suggest that
the $d$-wave is most likely to be the ground state of the system in
the $t$-$J$ model ($\alpha=1$). This conclusion is
 in agreement with many analytical and numerical
studies. Nevertheless, recent 32-site ED study showed that at $t/J>3.3$
the ground state of the two-hole $t$-$J$ model is a $p$-wave state with
total momentum  
${\bf P}=(\pi,\pi)$ \cite{Pakwo}. (In our case 
momenta ${\bf P}=(0,0)$ and ${\bf P}=(\pi,\pi)$ are equivalent
because of the existence of long-range order.) Furthermore, 
such a behavior is 
also suggested by recent series-expansion calculations \cite{Hamer}. 
Due to the 
apparently strong finite-size effects on the ED results,
it is difficult to conclude whether this behavior will survive
in the thermodynamic limit. Nevertheless, these new data show that
even in the pure $t$-$J$ model the final word on the problem of the
symmetry of the bound state is still to come.  

Besides effects arising from the $t$-$J$ model there are also concerns
about the robustness of the bound states in the real system. The
characteristic bound-state energy found in the $t$-$J$ model is of the
order of $-0.1J$ in the physical region of $t/J$. This energy is far
below the accuracy of the model itself. This means that in
order to describe the
real materials, the low-energy model must include at least additional
hopping terms ($t^\prime, t^{\prime\prime}$ etc.). 
It is worth noting again that because of these terms
the realistic dispersion is isotropic, which, as we have just shown,
can favor $p$-wave. More important conjecture can be made using the
results of Ref. \cite{BCS} and new ED data \cite{Pakwo1}. Since
the isotropic dispersion involves higher kinetic
energy, 
the overall pairing tendency is weakened by these additional
hopping terms. It can push all the bound states up to the continuum,
thus destroying the foundation of the preformed pair and phase
separation scenarios well before the physical region of parameters is
reached.  

\section{Conclusions}

In this paper we
attempt to address the issue of how generic $d$-wave pairing in
$t$-$J$-like models is and to clarify the problem of how this
symmetry is selected among others. We have thoroughly examined the  
$t$-$J_z$ model by means of analytical diagrammatic study supplemented by
our 32-site ED results. This formalism is tested
in the one-hole problem
and perfect agreement with ED results is found. The
ground state of two holes in the physical region $t/J>1$ is found 
to be a
$p$-wave bound state. From a careful analysis of the interactions
involved in the pairing problem we established that the pairing in the
$s$-wave channel is highly unfavorable because of the repulsive
character of the magnon-exchange interaction in this state. While
this reason is rather generic, the problem of selecting between the $p$-
and $d$-wave is much more subtle. In our analytical study these states
remain degenerate until the hole dispersion is included. In the $t$-$J_z$
model such a dispersion comes from higher order hopping
processes and is, in fact, very small. 
However, it has important consequences.
It is responsible not only for the significant
splitting between the $p$- and $d$-wave states, but also for the
destruction of
hole pairing at some critical $(t/J)|_c^{p,d}$. The isotropic form of the
dispersion favors $p$-wave as the ground state of the system. 

We have also considered the role of transverse fluctuations in the
perturbative limit $\alpha\ll 1$. They lead to an anisotropic hole
dispersion and an additional hole-hole interaction. We have shown that
both features favor $d$-wave as the lowest energy state. However, in
this case the selection between $p$- and $d$-wave strongly relies on the
anisotropic form of the hole dispersion,
and such a dispersion does not agree with the experimentally
measured dispersion in real systems.
Moreover, since the characteristic energy of the bound
states found in the $t$-$J$ models is small, we are skeptical that
these states can survive when additional kinetic energy terms are
included in the model. Our skepticism is supported by  earlier
studies \cite{BCS} and recent numerical data \cite{Tohyama,Pakwo1}.

Such conclusions for the pairing problem in $t$-$J$-like models
drawn from our study 
create a framework in which it might be necessary to look for 
$d$-wave superconductivity unrelated to the Bose-condensation of
preformed $d$-wave pairs. 
Furthermore, to make the phase separation scenario possible,
one might need to invoke arguments other than an {\it a priori}
existing  collapsing tendency in the spin-hole system.
 
\begin{center}
{\bf ACKNOWLEDGMENTS}
\end{center}

We are indebted to Prof. R. J. Gooding for suggesting this work
and for valuable discussions and comments.
We also thank F. Marsiglio for illuminative discussion, and
A. Castro Neto, M. Zhitomirsky, and O. Starykh for
comments. 
The work of A.L.C. was supported in part by the NSERC of Canada and 
the University of California and Los Alamos National Laboratory under 
the auspices of the US Department of Energy. 
The work of P.W.L. was supported by the Hong Kong RGC grant
HKUST6144/97P.
Numerical diagonalizations of the 32-site systems were performed
on the Intel Paragon at HKUST.


\begin{references}
\bibitem{comm1} The true assumption is 
weaker. One can require $\hat{\Gamma}(\epsilon)$ to be the sum of
a constant term and singular terms, latter having poles far from the
poles of the hole Green's functions in the integral. Approximating
$\hat{\Gamma}(\epsilon)$ by the constant term is valid provided that
$\Delta/2J\ll 1$.
\end{references}
\end{document}